\documentclass[fleqn,usenatbib]{mnras}

\usepackage{newtxtext,newtxmath}

\usepackage[T1]{fontenc}

\DeclareRobustCommand{\VAN}[3]{#2}
\let\VANthebibliography\thebibliography
\def\thebibliography{\DeclareRobustCommand{\VAN}[3]{##3}\VANthebibliography}

\usepackage{graphicx}	
\usepackage{amsmath}	
\usepackage{multirow}
\usepackage{afterpage}
\usepackage{comment}

\usepackage{ulem}

\title[The origin of QPOs in SMBH]{Radiation hydrodynamic simulations for the origin of quasi-periodic oscillations for accretion onto supermassive black holes}

\author[Yiyang Lin et al.]{
Yiyang Lin,$^{1,2}$
Erlin Qiao,$^{1,2}$\thanks{E-mail: qiaoel@nao.cas.cn}
Jifeng Liu,$^{1,2}$
Meng Guo$^{3,4}$
and Zikun Lin$^{5}$
\\
$^{1}$National Astronomical Observatories, Chinese Academy of Sciences, Beijing, 100101, China\\
$^{2}$School of Astronomy and Space Sciences, University of Chinese Academy of Sciences, 19A Yuquan Road, Beijing 100049, China\\
$^{3}$Key Laboratory of Computing Power Network and Information Security, Ministry of Education, Shandong Computer Science Center (National\\ 
Supercomputing Center in Jinan), Qilu University of Technology (Shandong Academy of Sciences), Jinan, Shandong 250013, China\\
$^{4}$Jinan Institute of Supercomputing Technology, Jinan, Shandong 250103, China\\
$^{5}$Department of Astronomy, Xiamen University, Xiamen, Fujian 361005, People’s Republic of China
}

\date{Accepted XXX. Received YYY; in original form ZZZ}

\pubyear{\the\year{}}

\begin{document}
\label{firstpage}
\pagerange{\pageref{firstpage}--\pageref{lastpage}}
\maketitle

\begin{abstract}
Quasi-periodic oscillation (QPO) has been detected in several accreting supermassive black hole (SMBH) systems, including active galactic nuclei (AGNs) and tidal disruption events (TDEs). However, despite that several models have been proposed, the physical origin of QPO is still unclear. In this paper, we performed radiation hydrodynamic simulations of accretion flow by injecting mass at a fixed radius, i.e. 10 Schwarzschild radius with different mass accretion rates, and setting the black hole (BH) mass to $10^7M_{\odot}$. We find that there are QPO signals by analyzing the mass inflow rates as a function of time from the simulations for different radii. The QPO frequencies from our simulations are well consistent with the radial epicyclic frequencies from analytic calculations for radius greater than a critical radius 3.8 Schwarzschild radius. This critical radius corresponds to the maximum epicyclic frequency, i.e. $\nu_{\rm r,max}$, in the radial direction. We proposed that $\nu_{\rm r,max}$ can be a good proxy for the observed QPO $\nu_{\rm QPO}$. Furthermore, assuming that our simulation results can be scaled to different BH masses $M_{\rm BH}$, we find that the theoretical relation of $\nu_{\rm r,max}$ as a function of $M_{\rm BH}$ can well match $\nu_{\rm QPO}$ as a function of $M_{\rm BH}$ for a sample of AGN and TDE. Finally, we discuss the effects of the BH mass, general relativity (GR), and other possible factors including the size of the mass injecting radius, viscosity and magnetic field on the simulation results.
\end{abstract}

\begin{keywords}
accretion, accretion discs -- black hole physics -- galaxies: active -- transients: tidal disruption events
\end{keywords}



\section{Introduction}

Quasi-periodic oscillation (QPO) has been detected in several supermassive black hole (SMBH) systems, containing active galactic nuclei (AGNs) RE J1034+396 \citep{Gierlinski2008, Alston2014} and 1ES 1927+654 \citep{Masterson2025}, tidal disruption events (TDEs) ASASSN-14li \citep{Pasham2019}, Swift J164449.3+573451 \citep{Reis2012} and TDE candidate 2XMM J123103.2+110648 \citep{Lin2013, lin2017, Webbe2023}. These QPOs in SMBH systems show a period range of several minutes to hours.

Theoretical attempts to explain the QPO phenomenon in accreting black hole (BH) systems have been made for decades.
\citet{Kluzniak&Abramowicz2001} proposed the importance of the resonance between the orbital frequencies and the epicyclic frequencies of accretion disks for explaining the observed 3:2 frequency in black hole X-ray binaries (BHXBs). This observed 3:2 frequency is classified as high frequency QPO (HFQPO) \citep[e.g.][for review]{Remillard2006ARA&A}.
Thereafter, several numerical simulations have continued to study this idea.
\citet{Sramkova2007} and \citet{Parthasarathy2016} showed QPO signatures in perturbed tori by pseudo-Newtonian (PN) hydrodynamic (HD) simulations. \citet{Neill2009} found trapped global oscillation modes in their two-dimensional (2D) PN-HD simulations of geometrically thin disks with $\alpha$-model viscosity, expecting to evaluate the potential role of these oscillation modes in producing astronomically observed HFQPOs in accreting BH binary systems. \citet{Kato2004} presented a twin HFQPO signature in their three-dimensional (3D) PN magnetohydrodynamic (MHD) simulations. \citet{Schnittman2006} identified transient QPO pairs by ray tracing a 3D general relativistic magnetohydrodynamic (GRMHD) simulation of an accreting Schwarzschild BH. \citet{Mishra2019, Mishra2020} generated QPO signatures using their 2D viscous GR radiation hydrodynamic (RHD) simulations in accretion disks on both the gas- and radiation-pressure-dominated branches around stellar-mass BHs. 
Tilted disks have also been studied. 
In \citet{Fragile2008} and \citet{Henisey2009}, the authors analyzed GRMHD simulations of tilted accretion disks around Kerr BHs. \citet{Musoke2023} present QPO signatures by their 3D GRMHD simulations of a highly tilted ($65^{\circ}$) thin accretion disk around a rapidly spinning stellar-mass BH, demonstrate that disk tearing naturally produces both low-frequency (precession-driven) and high-frequency (epicyclic) QPOs, offering a unified mechanism for observed QPOs in X-ray binaries.



However, despite that a number of simulations have studied QPO signatures,  very fewer simulations focus on QPOs in SMBHs.
In \citet{Chakrabarti2004}, the authors performed 2D axisymmetric smoothed particle hydrodynamics (SPH) simulations for the properties of accretion flow around a SMBH with PN potential adopted. In the simulations, the flow is assumed to be sub-Keplerian and inviscid, and the radiative cooling is considered. Their simulations produced QPO-like signals with characteristic timescales ranging from hours to weeks, which are roughly comparable to the observed QPOs in SMBHs.


More recently, \citet{Dihingia2025} performed 3D GRMHD simulations of low-angular-momentum accretion flows in Kerr spacetime.
They show that cHz QPO signals can be found in the innermost regions of magnetized accretion flows for a SMBH of $10^7M_{\odot}$.
It is clear that GRMHD simulations can capture more detailed dynamics of accretion flow compared with earlier SPH simulations. However, 
the simulation in \citet{Chakrabarti2004} can generate QPO signals closer 
to the observations. As for the treatment for the radiation, in \citet{Chakrabarti2004} only radiative cooling is considered in the simulations. While in \citet{Dihingia2025}, the treatment for the radiative transfer is separate with the GRMHD simulations. Meanwhile, in both \citet{Chakrabarti2004} and \citet{Dihingia2025}, the sub-Keplerian (low angular momentum) accretion flows are assumed in the simulations, which are different from the focus in the present paper.


In this work, we perform radiation hydrodynamic simulations of accretion flows around a SMBH using the Athena++ code with the most advanced treatment for radiative transfer \citep{2021ApJS..253...49J,2022ApJS..263....4J}. 
Compared with previous studies, our simulation self-consistently couples the dynamics of the accretion flow with radiative transfer, allowing us to better investigate the properties of the accretion flow and the related features.
In Section \ref{sec:2} we introduce the simulation setup. In Section \ref{sec:3} we present our results, including the evolution of accretion flows, the 
emergence of QPO signals, theoretical analysis and the comparison with observations. Discussions are given in Section \ref{sec:4}.

\section{Simulation Setup} \label{sec:2}
We use the Athena++ code \citep{Stone2020,Athena21} to perform two-dimensional axisymmetric viscous radiation hydrodynamic (RHD) simulations of accretion flows around a SMBH. In the simulation, we adopt spherical coordinates ($r$, $\theta$, $\phi$). The equations of hydrodynamics coupled with the time-dependent radiative transfer equations we solved can refer to equation (1) and (2) (combining equations (3) and (5)) in \citet{Qiao2025}. In this paper, we fix the mass of BH $M_{\rm BH}=10^7M_{\odot}$. We only consider the $r\phi$ component of the stress tensor and fix the viscosity parameter $\alpha=0.1$ as that of \citet{Qiao2025}. We carried out three groups of simulations by continuously injecting mass at mass injecting radius $R_{\rm inject} = 10R_{\rm S}$ (with $R_{\rm S} \equiv 2GM_{\rm BH}/c^2$ being the Schwarzschild radius) with the mass injection rate $\dot M_{\rm inject}=0.01\dot M_{\rm Edd}$, $0.1\dot M_{\rm Edd}$ and $2\dot M_{\rm Edd}$ (with $\dot M_{\rm Edd}$ = $1.39 \times 10^{18} M_{\rm BH}/M_{\rm \odot} \rm \ g s^{-1}$), respectively. 
The injection rates are set to be constant in the simulations. We assume that the rotational velocity of the injected gas equals the local Keplerian velocity.

Specifically, in the simulation, we inject gas in the region $9R_{\rm S} \leqslant r \leqslant 11R_{\rm S}$ and $\frac{19}{40}\pi \leqslant \theta \leqslant \frac{21}{40}\pi$. Our simulation has a computational domain with $2R_{\rm S} \leqslant r \leqslant 10^5 R_{\rm S}$ in the radial direction and $0 \leqslant \theta \leqslant \pi$ in the $\theta$ direction with resolution $N_r \times N_{\theta} = 768 \times 256$. Specifically, in the region $2R_{\rm S} \leqslant r \leqslant 3R_{\rm S}$, we set 16 uniformly spaced cells. We set 240 logarithmically spaced cells in the region $3R_{\rm S} \leqslant r \leqslant 50R_{\rm S}$. In the region $50R_{\rm S} \leqslant r \leqslant 200R_{\rm S}$, we set 128 logarithmically spaced cells, and in the region $200R_{\rm Ss} \leqslant r \leqslant 93405R_{\rm S}$, we set 380 logarithmically spaced cells. Finally, in the region $93405R_{\rm S} \leqslant r \leqslant 10^5 R_{\rm S}$, we set 4 uniformly spaced cells. In the $\theta$ direction, we set 256 non-uniformly spaced cells. The size of the cells is set to be smaller close to the mid-plane and the distribution of the cells is symmetric to the mid-plane. Outflow boundary conditions are considered at inner and outer radial boundary, while at $\theta=0$ and at $\theta = \pi$ we use the reflecting boundary conditions.
The pseudo-Newtonian potential is used to mimic the effects of general relativity around a Schwarzschild BH, i.e.,

\makeatletter
\@fleqnfalse
\makeatother
\begin{equation}
\centering
\phi=-\frac{GM_{\rm BH}}{r-R_{\rm S}},
\label{eq:1}
\end{equation}
\makeatletter
\@fleqntrue
\makeatother

\noindent where $G$ is the gravitational constant and $r$ is the distance to the central BH. 
At the inner and outer radial boundaries, we take outflow boundary conditions, while at $\theta =0$ and at $\theta = \pi$ we take reflecting boundary conditions. In this paper, the simulation is running until $t=400$ day since mass injection, corresponding to approximately 1394 orbital periods at $R_{\rm inject} = 10R_{\rm S}$, ensuring that the system can evolve enough times to capture the system dynamics.

\section{Results} \label{sec:3}
\subsection{Evolution of the accretion flow configuration for different $\dot M_{\rm inject}$} \label{subsec:3.1}
Mass accretion rate is believed to be one of the most important
parameters which can control the dynamics and the radiation of the accretion flow around BHs \citep[][for review]{Abramowicz2013LRR....16....1A}. In order to explore the behavior of QPOs with mass accretion rate, in this paper, we consider three different mass injection rates, i.e., $\dot{M}_{\rm inject}=0.01\dot{M}_{\rm Edd}$, $0.1\dot{M}_{\rm Edd}$, and $2\dot{M}_{\rm Edd}$ for the simulations respectively. 
Simply speaking, for $\dot{M}_{\rm inject}=0.01\dot{M}_{\rm Edd}$ and $0.1\dot{M}_{\rm Edd}$, the accretion flow fall in the geometrically thin, optically thick disk regime \citep[][]{Shakura1973A&A....24..337S},
while for $\dot{M}_{\rm inject}=2\dot{M}_{\rm Edd}$, the accretion flow fall in the geometrically thick, optically thick slim disk regime \citep[][]{Abramowicz1988ApJ...332..646A}.

\begin{figure*}
    \centering
    \includegraphics[width=0.24\textwidth]{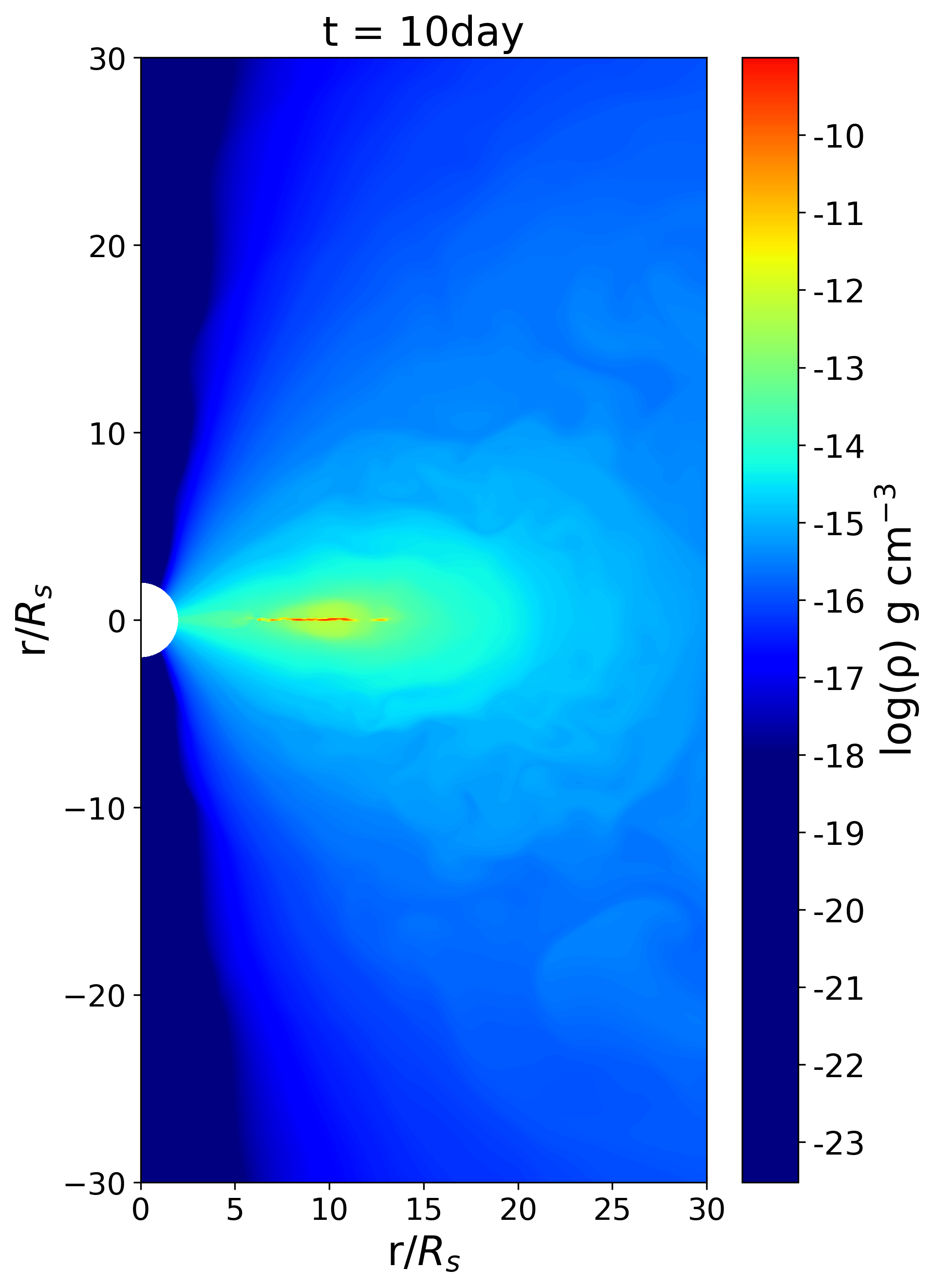}
    \hspace{0.01\textwidth}
    \includegraphics[width=0.24\textwidth]{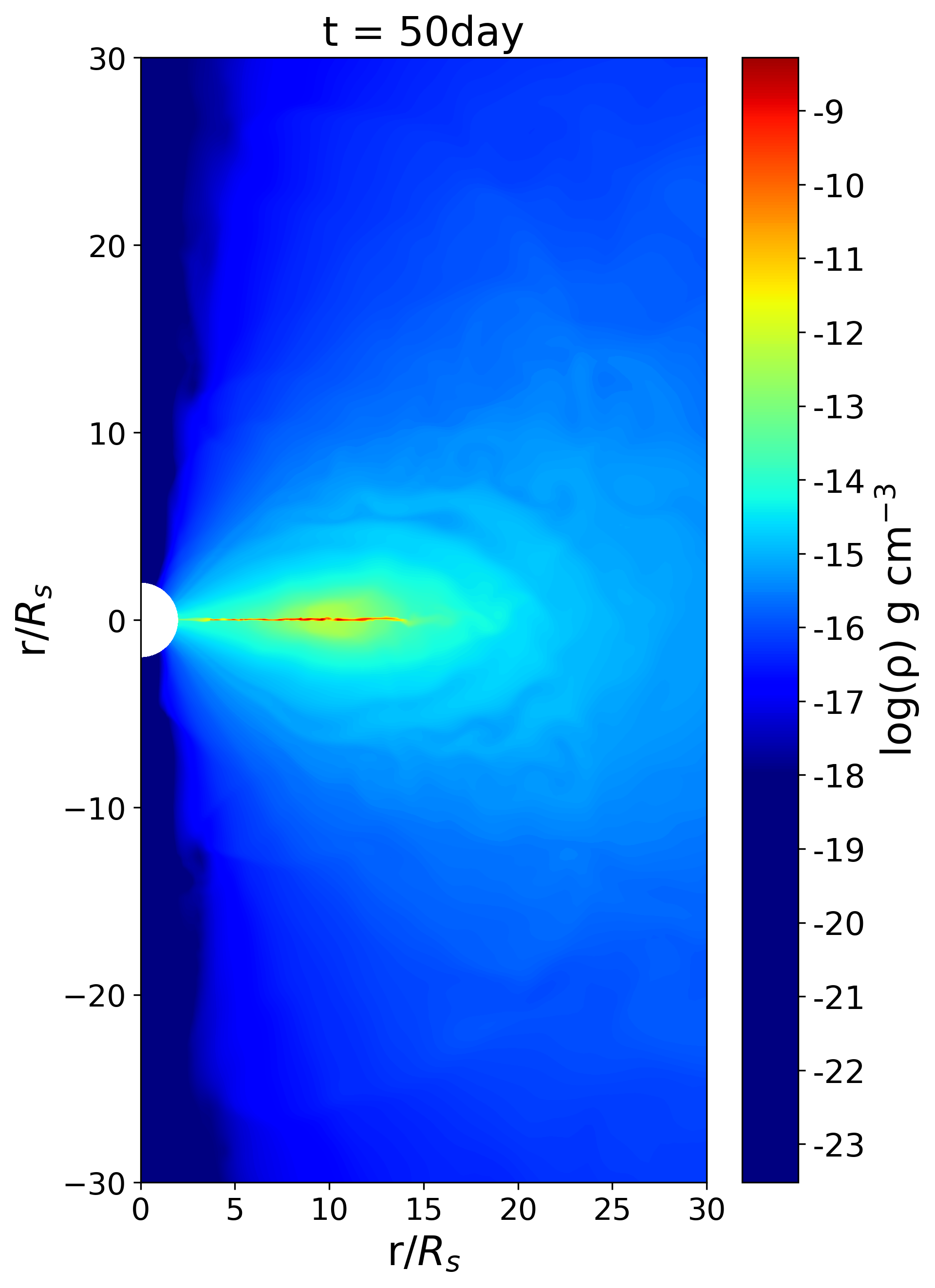}
    \hspace{0.01\textwidth}
    \includegraphics[width=0.24\textwidth]{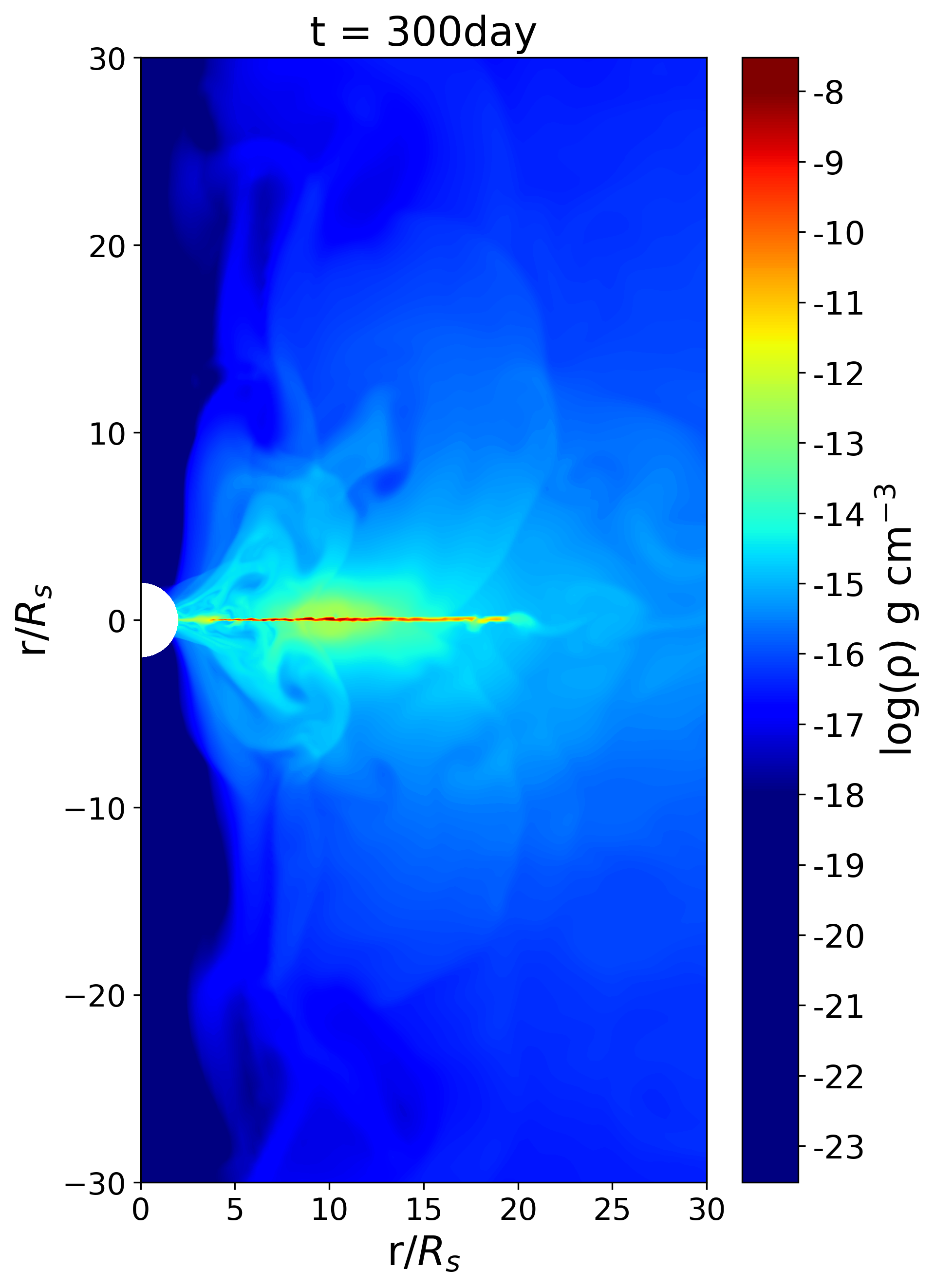}
    \caption{Snapshots of gas density (in $r-\theta$ plane) at $t=10,50,300$ day since the injection of matter at $R_{\rm inject}$ with $\dot M_{\rm inject} = 0.01\dot M_{\rm Edd}$.}
    \label{fig:1}
\end{figure*}

In Fig. \ref{fig:1}, we plot snapshots of the gas density at $t=10, 50, 300$ day since injection of matter at $R_{\rm inject}$ with $\dot M_{\rm inject} = 0.01\dot M_{\rm Edd}$.
It can be seen that at $t=10$ day, the main body of the accreted mass, i.e. the red part in the equatorial plane in the left panel of Fig. \ref{fig:1}, is filling the gap between $R_{\rm inject}$ and the BH. At this time, a small fraction of the accreted mass has diffused onto the BH, and conversely, another small fraction of the accreted mass is blown away outward in the form of the outflow. One can see the diffused cyan part beyond $R_{\rm inject}$ in the left panel of Fig. \ref{fig:1} for details. At $t=50$ day, the main body of the accreted mass in the equatorial plane has filled the gap between $10R_{\rm S}$ and the BH, and stable accretion is ongoing. Meanwhile, at this time, the main body of the accreted mass in the equatorial plane diffuses outward significantly due to angular momentum conservation. One can see the middle panel of Fig. \ref{fig:1} for details. At $t=300$ day, the global properties of the accretion flows are similar to those of at $t=50$ day except that the main body of the accreted mass in the equatorial plane diffuses further outward. One can see the right panel of Fig. \ref{fig:1} for clarity.

\begin{figure*}
    \centering
    \includegraphics[width=0.24\textwidth]{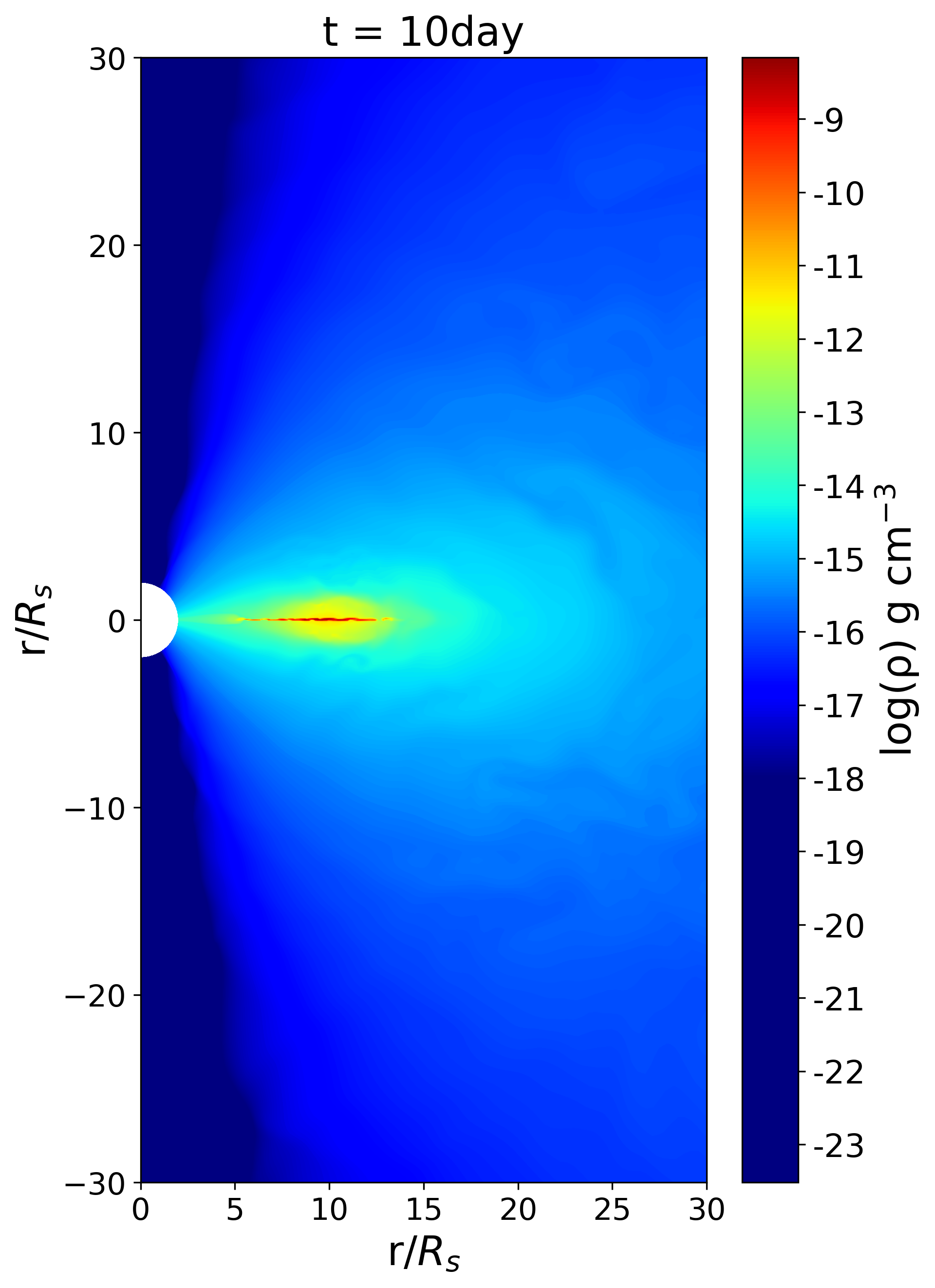}
    \hspace{0.01\textwidth}
    \includegraphics[width=0.24\textwidth]{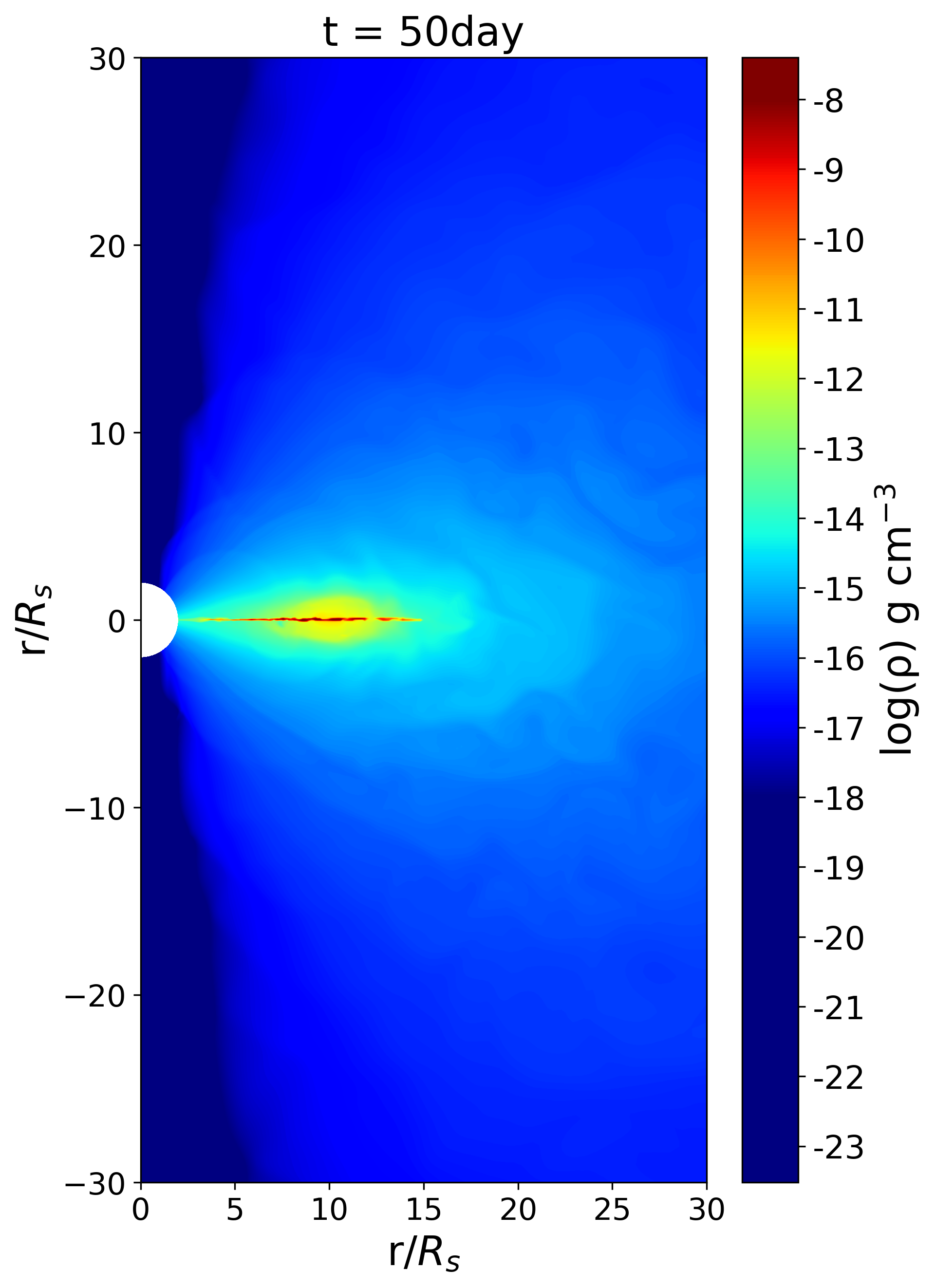}
    \hspace{0.01\textwidth}
    \includegraphics[width=0.24\textwidth]{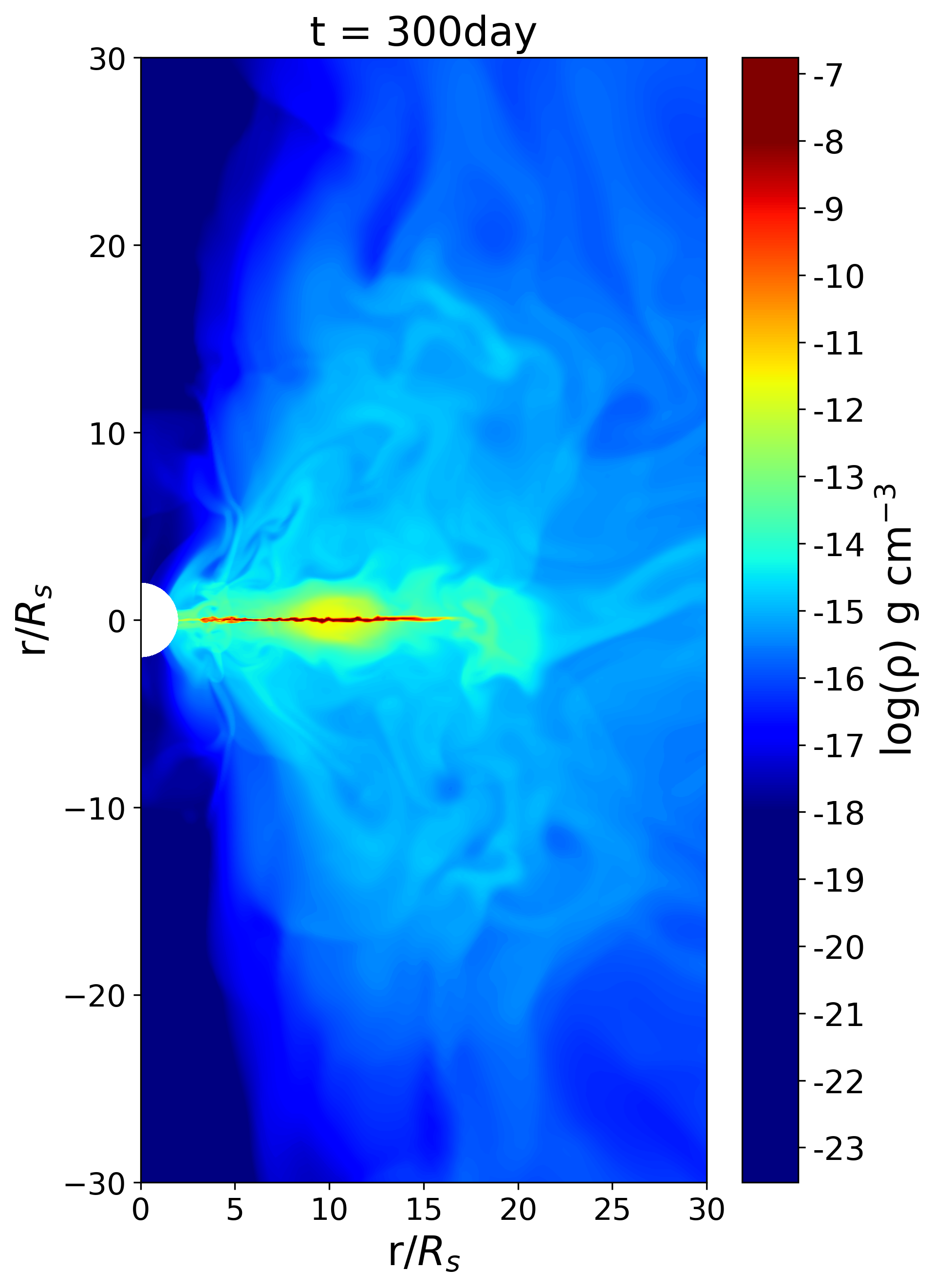}
    \caption{Snapshots of gas density (in $r-\theta$ plane) at $t=10,50,300$ day since the injection of matter at $R_{\rm inject}$ with $\dot M_{\rm inject} = 0.1\dot M_{\rm Edd}$.}
    \label{fig:2}
\end{figure*}

In Fig. \ref{fig:2}, we plot snapshots of the gas density at $t=10, 50, 300$ day since injection of matter at $R_{\rm inject}$ with $\dot M_{\rm inject} = 0.1\dot M_{\rm Edd}$. The global properties of the accretion flows are similar to that of the case with $\dot M_{\rm inject} = 0.01\dot M_{\rm Edd}$ expect that the global densities of the accretion flow increase. 

\begin{figure*}
    \centering
    \includegraphics[width=0.24\textwidth]{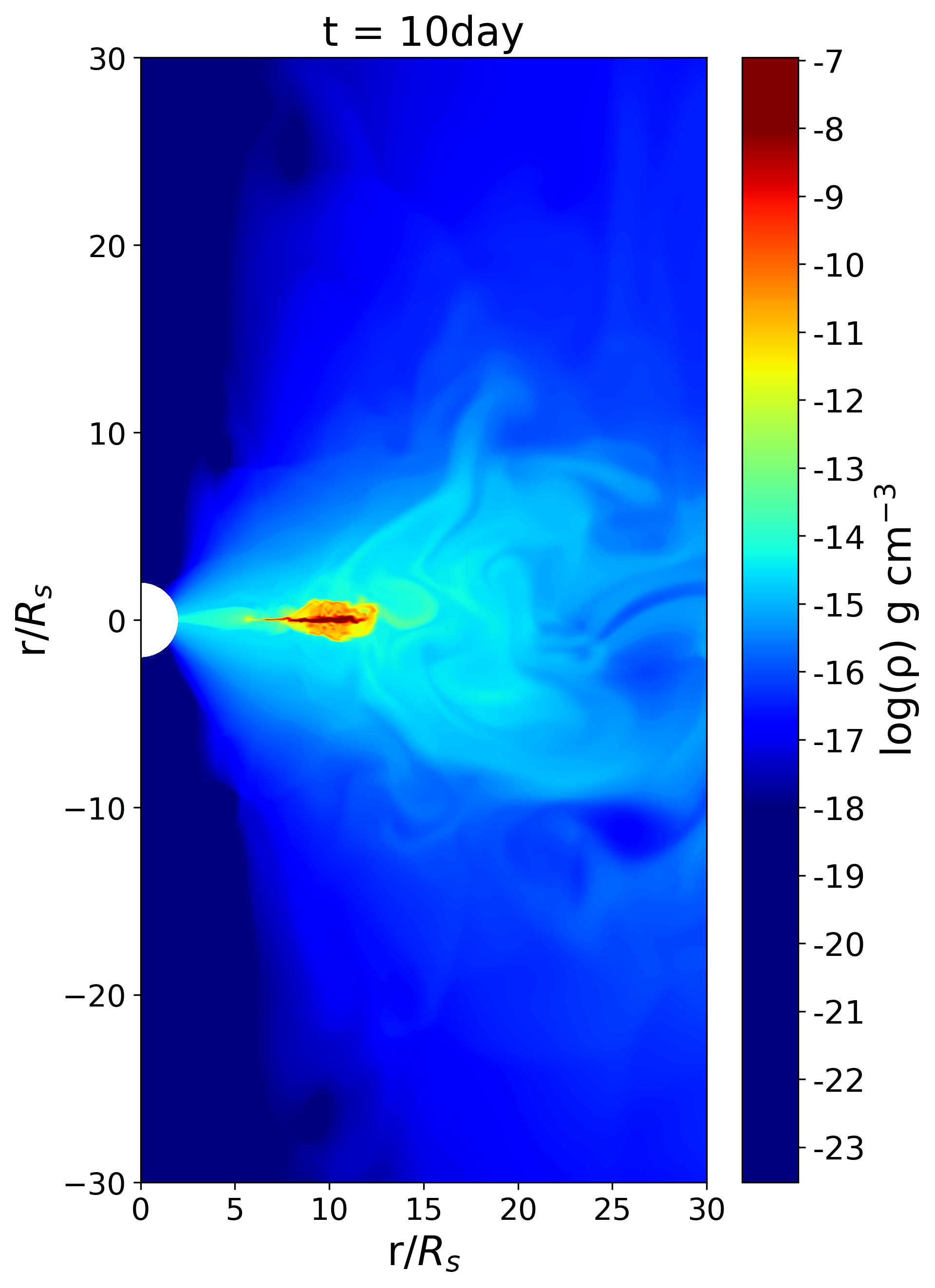}
    \hspace{0.001\textwidth}
    \includegraphics[width=0.24\textwidth]{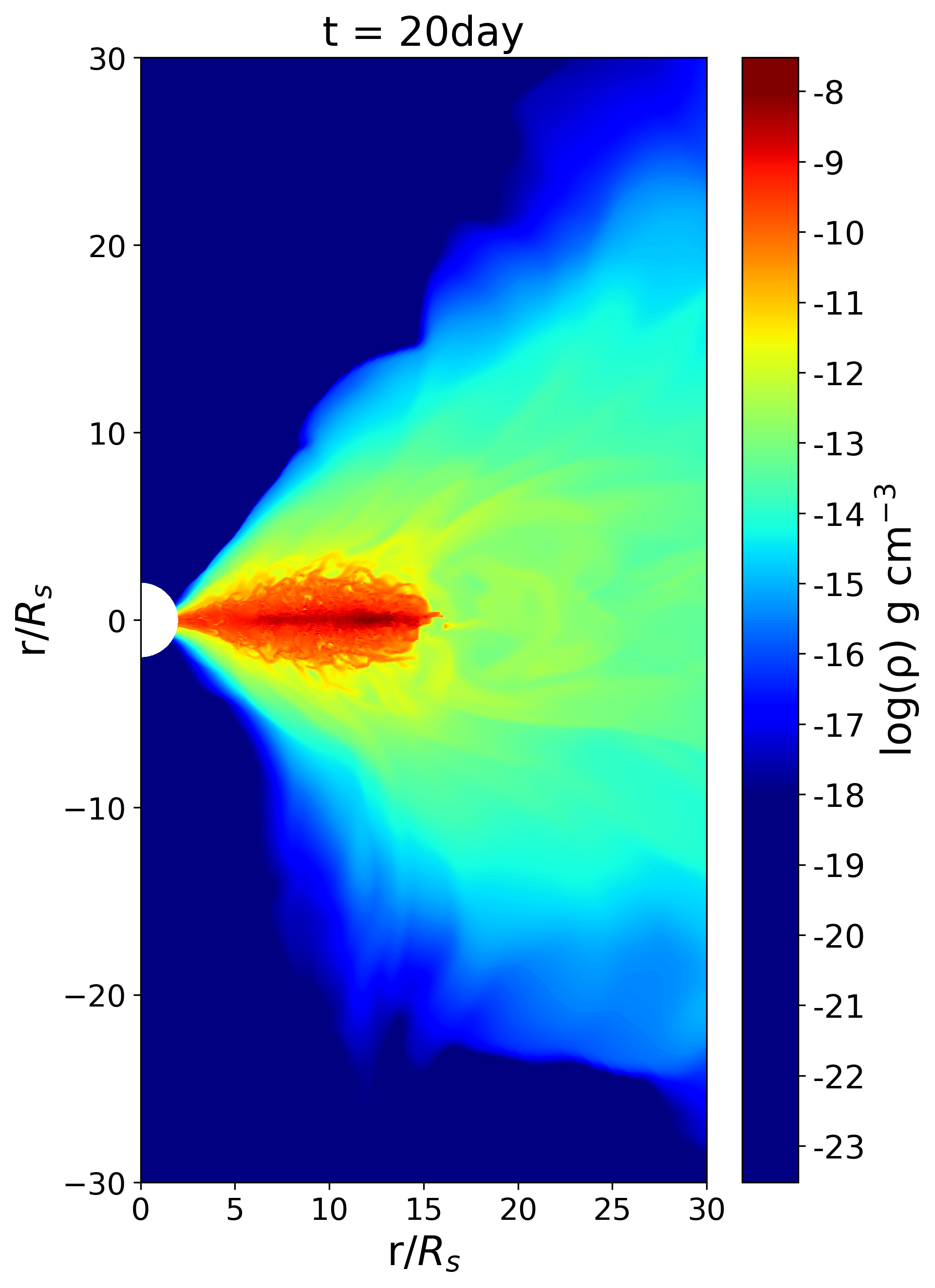}
    \hspace{0.001\textwidth}
    \includegraphics[width=0.24\textwidth]{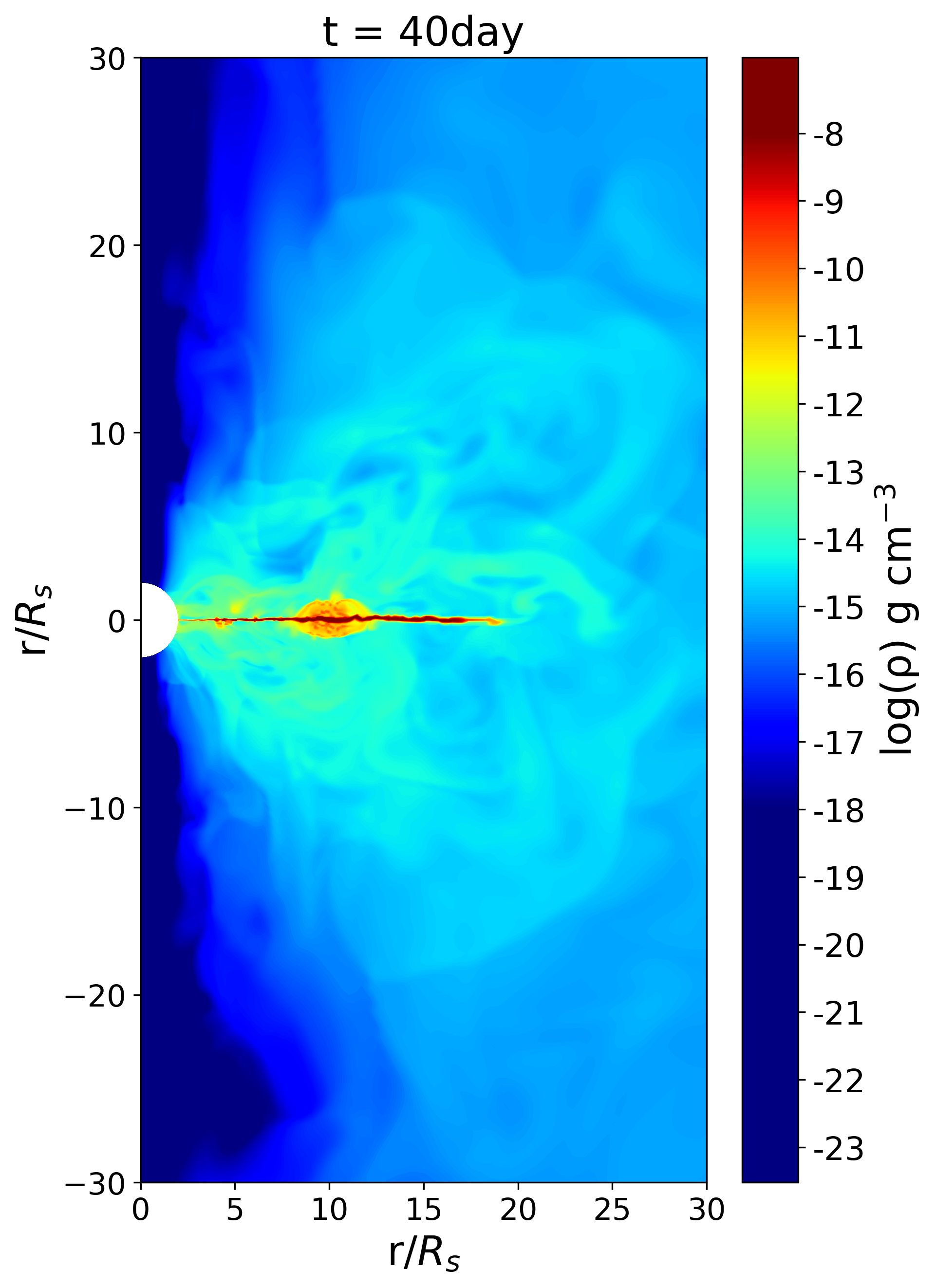}
    \hspace{0.001\textwidth}
    \includegraphics[width=0.24\textwidth]{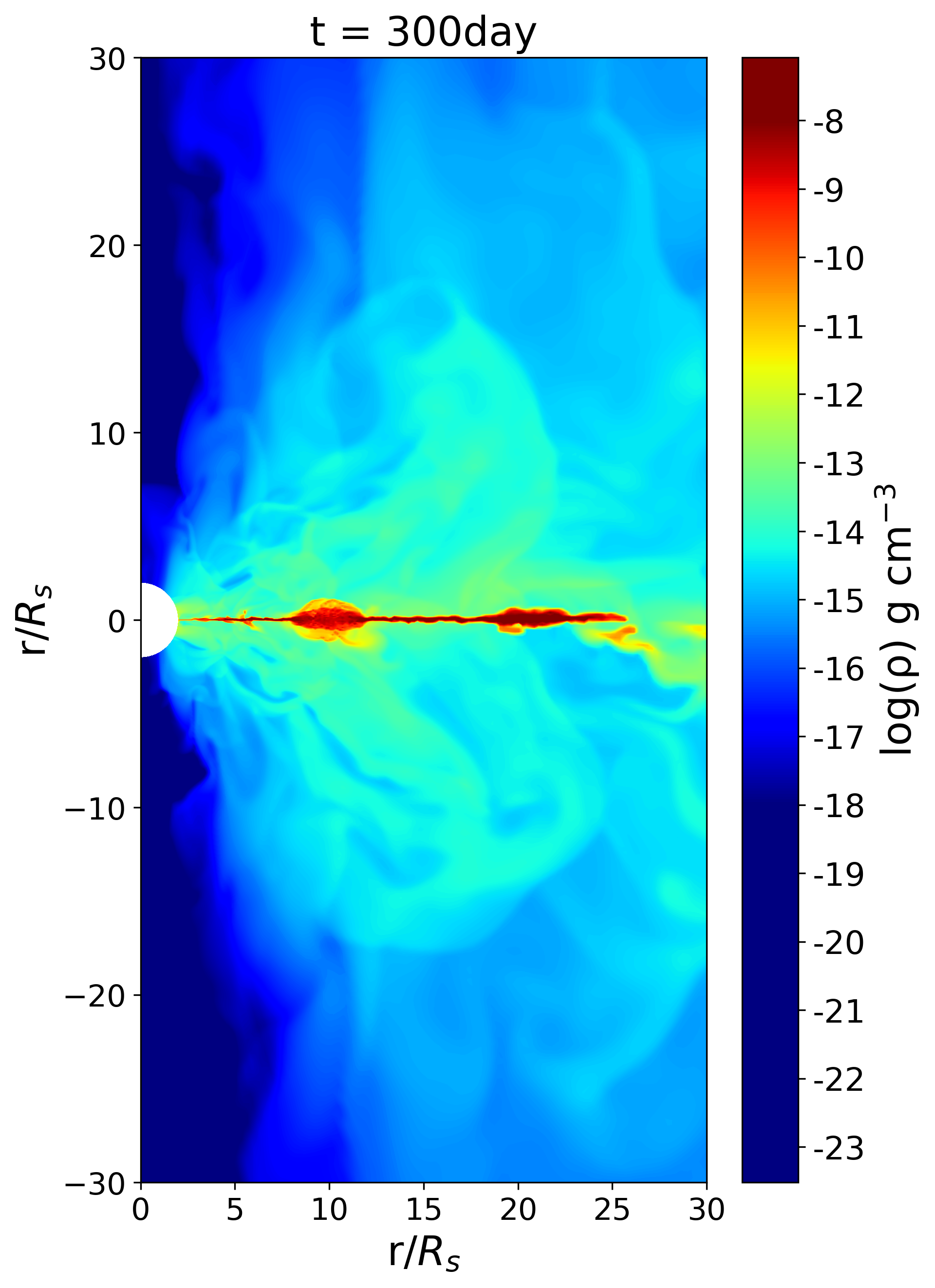}
    \caption{Snapshots of gas density (in $r-\theta$ plane) at $t=10,20,40,300$ day since the injection of matter at $R_{\rm inject}$ with $\dot M_{\rm inject} = 2\dot M_{\rm Edd}$.}
    \label{fig:3}
\end{figure*}

In Fig. \ref{fig:3}, we plot snapshots of the gas density at $t=10, 20, 40, 300$ day since injection of matter at $R_{\rm inject}$ with $\dot M_{\rm inject} = 2\dot M_{\rm Edd}$. Since the significant increase of the mass injection rate, the global properties of the accretion flow are significantly different from that of the cases with $\dot M_{\rm inject} = 0.01\dot M_{\rm Edd}$ and $\dot M_{\rm inject} = 0.1\dot M_{\rm Edd}$. There are two key different points: (1) The global densities of the accretion flow (including the inflow and the outflow) increase. Meanwhile, the main body of the accretion flow becomes geometrically thick in the vertical direction compared with that the main body of the accretion flow being constrained in the very thin equatorial plane for $\dot M_{\rm inject} = 0.01\dot M_{\rm Edd}$ and $\dot M_{\rm inject} = 0.1\dot M_{\rm Edd}$. (2) There are significant state transitions between high state and low state. For example, at $t=20$ day, the accretion is in a high state. At this time the main body of the accretion flows are extended in the vertical direction, and there are strong outflows. While at $t=40$ day, the accretion is in low state. At this time the main body of the accretion flows is constrained in a very narrow region in the equatorial plane, and there are very weak outflows. 
This state transitions last until the end of the simulation in this paper of $t=400$ day. As an example, we plot the snapshot of the gas density at $t=300$ day. At this time, the accretion is in low state, one can see the fourth panel in Fig. \ref{fig:3} for details. 
The state transition in this simulation is suggested to be the radiation pressure instability as predicted by early theoretical analyses and one-dimensional calculation of the time-dependent accretion disk for the mass accretion rate around $\dot M_{\rm Edd}$ \citep{Nayakshin2000,Janiuk2002,Shen2014,Lu2022,Piro2025,Guo2026}. 

\subsection{Evolution of $\dot M_{\rm in}$ and the emergence of QPO} \label{subsec:3.2}

\subsubsection{The case for $\dot M_{\rm inject}=0.01 \dot M_{\rm Edd}$ and $\dot M_{\rm inject}=0.1 \dot M_{\rm Edd}$ } \label{subsubsec:3.2.1}

For a BH accreting system, the mass inflow rate is one of the most important physical quantities that can reflect the global properties of the accretion flow. We define the mass inflow rate $\dot M_{\rm in}$ as follows,

\makeatletter
\@fleqnfalse
\makeatother
\begin{equation}
\dot M_{\rm in}(r)=2\pi {r}^2 \int_{-\pi /2}^{\pi /2} \rho {\rm min}(v_{\rm r},0) {\rm sin}\theta d\theta.
\label{eq:2}
\end{equation}
\makeatletter
\@fleqntrue
\makeatother

\begin{figure*}
    \centering
    \includegraphics[width=1\textwidth]{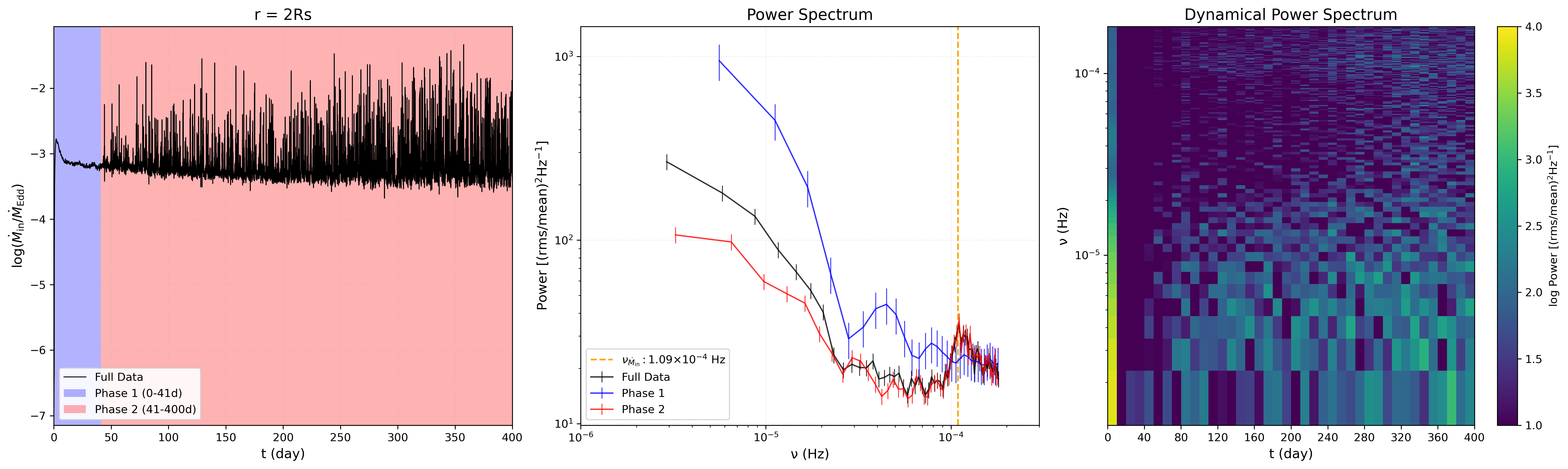}
    \hfill
    \includegraphics[width=1\textwidth]{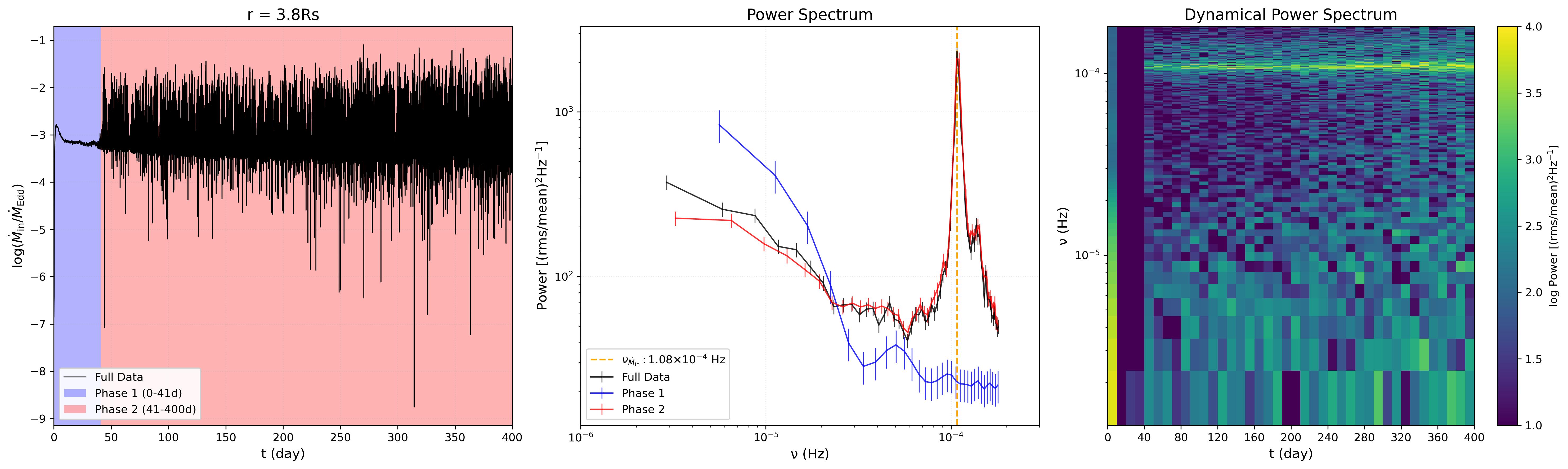}
    \caption{Evolution of the scaled mass inflow rate $\dot M_{\rm in}/\dot M_{\rm Edd}$, the corresponding power spectrum and the dynamical power spectrum with mass injection rate $\dot M_{\rm inject} = 0.01\dot M_{\rm Edd}$. 
    Left panel: $\dot M_{\rm in}/\dot M_{\rm Edd}$ as a function of $t$ at $r=2R_{\rm S}$ (top) and $r=3.8R_{\rm S}$ (bottom) respectively. The purple region and the red region refer to phase 1 ($0-41$) day and phase 2 ($41-400$) day respectively. Middle panel: Power spectrum calculated with FFT for the full data, phase I data and phase 2 data at $r=2R_{\rm S}$ (top) and $r=3.8R_{\rm S}$ (bottom) respectively. Right panel: Dynamical power spectrum calculated with FFT at $r=2R_{\rm S}$ (top) and $r=3.8R_{\rm S}$ (bottom) respectively.}
    \label{fig:4}
\end{figure*}

In the left panel of Fig. \ref{fig:4}, we plot the scaled mass inflow rates $\dot M_{\rm in}/\dot M_{\rm Edd}$ as a function of time $t$ at $r=2R_{\rm S}$ (top) and $r=3.8R_{\rm S}$ (bottom), respectively, for $\dot M_{\rm inject} = 0.01\dot M_{\rm Edd}$. 
We choose $r=2R_{\rm S}$ because it corresponds to the inner boundary of our simulation domain, while $r=3.8R_{\rm S}$ is close to the radius $r=(2+\sqrt{3})R_{\rm S}$ where the theoretical radial epicyclic frequency reaches its maximum in the adopted PN potential (please refer to Section. \ref{subsec:3.3} for details).
We divide $\dot M_{\rm in}/\dot M_{\rm Edd}$ as a function of time $t$ into two parts, i.e., phase 1 ($0-41$) day and phase 2 ($41-400$) day, which correspond to the phase during which the accreted mass is filling the gap between the mass injection radius and the BH, and the phase during which the stable accretion is ongoing, respectively.
With the time series of $\dot M_{\rm in}/\dot M_{\rm Edd}$, we calculate the power spectrum and the dynamical power spectrum (in the practical calculation, ${\rm log} (\dot M_{\rm in}/\dot M_{\rm Edd})$ used for calculating the power spectrum and the dynamical power spectrum.) with the method of Fast Fourier Transform (FFT; \citealt{Cooley&Tukey1965}) using the \textit{Lightcurve}, \textit{AveragedPowerspectrum} and \textit{DynamicalPowerspectrum} packages in \textit{Stingray} \citep{Huppenkothen2019}. The QPO frequencies for different radii are given in Table \ref{tab:1} and FFT parameters are given in the note.

One can see the middle panel of Fig. \ref{fig:4} for details. Specifically, the blue, red and the black symbols refer to the power spectrum calculated with phase 1 data ($0-41$) day,  phase 2 ($41-400$) day and the full data ($0-400$) day respectively. It is clear there is no QPO found if the phase 1 data are used for both at $r=2R_{\rm S}$ and $r=3.8R_{\rm S}$. However, if the phase 2 data and full data are used, there is QPO. Specifically, the QPO frequency $\nu_{\dot M_{\rm in}}$ is $1.09\times 10^{-4}$ Hz for both the phase 2 data and full data at $r=2R_{\rm S}$. The power of QPO is weak at $r=2R_{\rm S}$, which can be reflected in the dynamical power spectrum in the upper right panel of Fig. \ref{fig:4}. As for $r=3.8R_{\rm S}$, the QPO frequency $\nu_{\dot M_{\rm in}}$ is $1.08\times 10^{-4}$ Hz for both phase 2 data and full data. At $r=3.8R_{\rm S}$, the QPO frequency of $1.08\times 10^{-4}$ Hz is very strong, which can be clearly seen in the dynamical power spectrum in the lower right panel of Fig. \ref{fig:4}. In the calculation of the dynamical power spectrum, the time bin is chosen to be $10$ days.

\begin{figure*}
    \centering
    \includegraphics[width=1\textwidth]{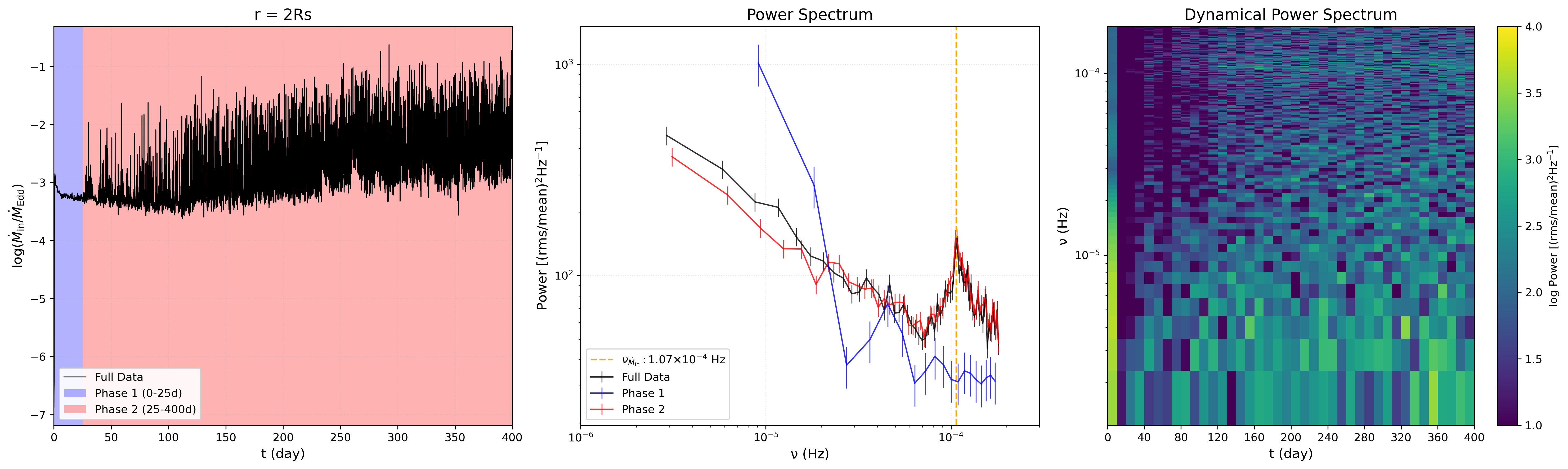}
    \hfill
    \includegraphics[width=1\textwidth]{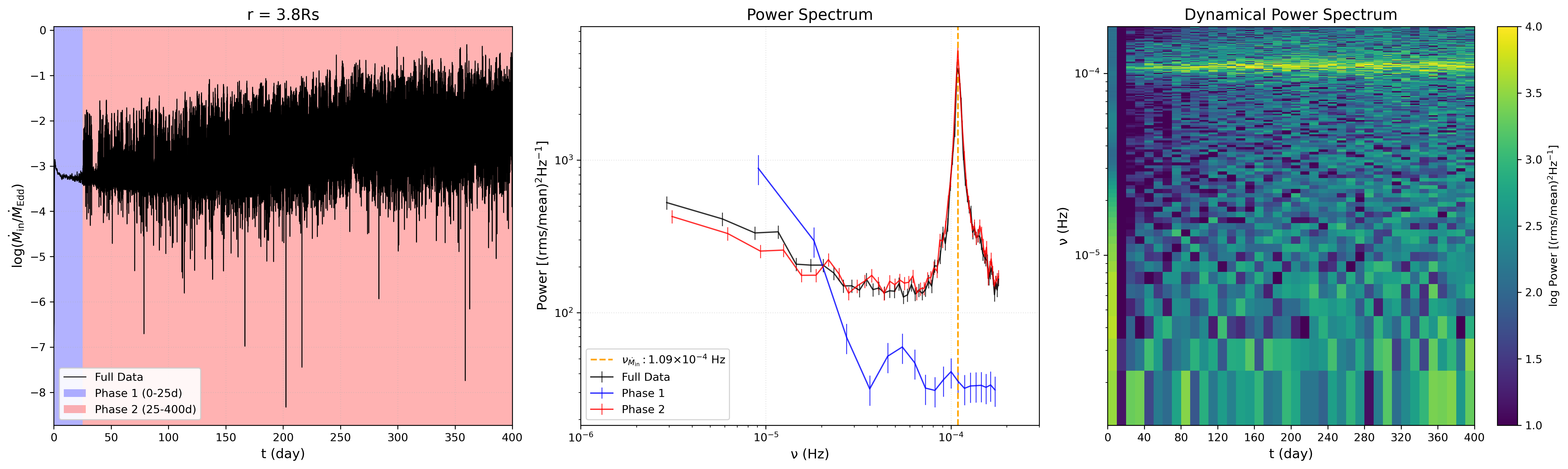}
    \caption{Evolution of the scaled mass inflow rate $\dot M_{\rm in}/\dot M_{\rm Edd}$, the corresponding power spectrum and the dynamical power spectrum with mass injection rate $\dot M_{\rm inject} = 0.1\dot M_{\rm Edd}$. 
    Left panel: $\dot M_{\rm in}/\dot M_{\rm Edd}$ as a function of $t$ at $r=2R_{\rm S}$ (top) and $r=3.8R_{\rm S}$ (bottom) respectively. The purple region and the red region refer to phase 1 ($0-25$) day and phase 2 ($25-400$) day respectively. Middle panel: Power spectrum calculated with FFT for the full data, phase I data and phase 2 data at $r=2R_{\rm S}$ (top) and $r=3.8R_{\rm S}$ (bottom) respectively. Right panel: Dynamical power spectrum calculated with FFT at $r=2R_{\rm S}$ (top) and $r=3.8R_{\rm S}$ (bottom) respectively.}
    \label{fig:5}
\end{figure*}

The scaled mass inflow rates $\dot M_{\rm in}/\dot M_{\rm Edd}$ as a function of $t$, the power spectrum and the dynamical power spectrum for $r=2R_{\rm S}$ (top) and $r=3.8R_{\rm S}$ (bottom) for $\dot M_{\rm inject} = 0.1\dot M_{\rm Edd}$ are plotted in Fig. \ref{fig:5}, which are similar to that of $\dot M_{\rm inject} = 0.01\dot M_{\rm Edd}$. We also divide $\dot M_{\rm in}/\dot M_{\rm Edd}$ as a function of time $t$ into two parts, i.e. phase 1 ($0-25$) day and phase 2 ($25-400$) day, which correspond to the phase that the accreted mass is filling the gap between the mass injection radius and the BH, and the phase that the stable accretion is ongoing, respectively. By doing FFT with phase 1 data ($0-25$) day, there is no QPO found for both at $r=2R_{\rm S}$ and $r=3.8R_{\rm S}$. However, if doing FFT with phase 2 data ($25-400$) day and the full data ($0-400$) day, there is QPO found. Specifically, the QPO frequency $\nu_{\dot M_{\rm in}}$ is $1.07\times 10^{-4}$ Hz for both the phase 2 data and full data at $r=2R_{\rm S}$, while the QPO frequency $\nu_{\dot M_{\rm in}}$ is $1.09\times 10^{-4}$ Hz for both the phase 2 data and full data at $r=3.8 R_{\rm S}$. One can see the blue, red and the black symbols in the middle panel of Fig. \ref{fig:5} for details. The dynamical power spectrum is plotted in the upper right panel of Fig. \ref{fig:5} for $r=2R_{\rm S}$ and in the lower right panel of Fig. \ref{fig:5} for $r=3.8R_{\rm S}$, respectively. In the calculation, the time bin is also chosen to be 10 days. The QPO frequencies for different radii are given in Table \ref{tab:2} and FFT parameters are given in the note.

\subsubsection{The case for $\dot M_{\rm inject}=2 \dot M_{\rm Edd}$ }
\label{subsubsec:3.2.2}

\begin{figure*}
    \centering
    \includegraphics[width=1\textwidth]{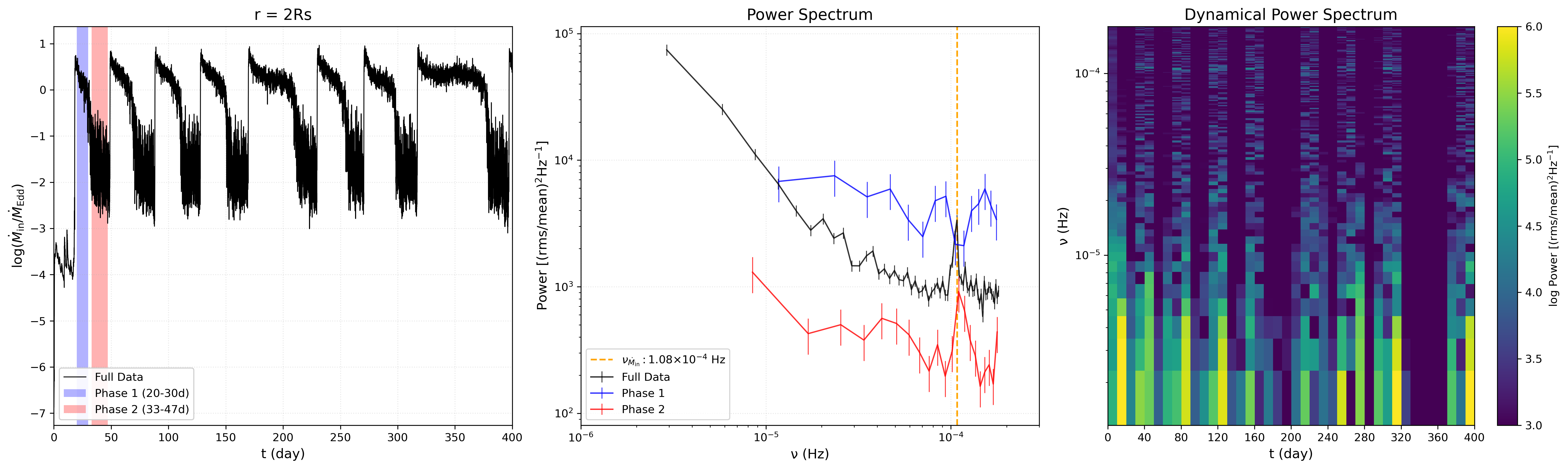}
    \hfill
    \includegraphics[width=1\textwidth]{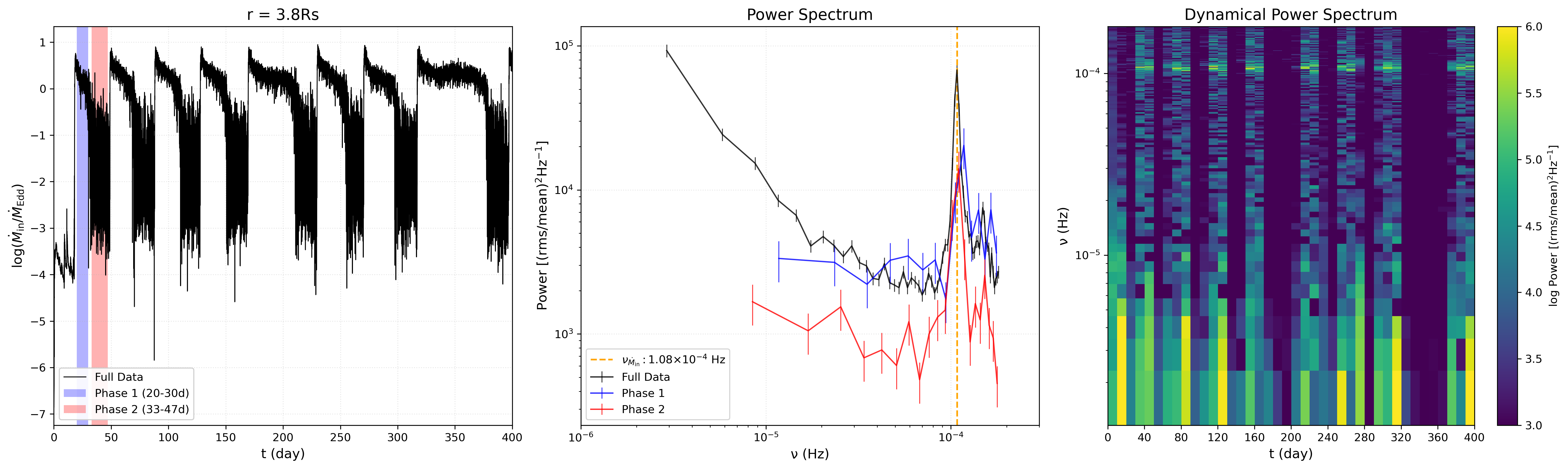}
    \caption{Evolution of the scaled mass inflow rate $\dot M_{\rm in}/\dot M_{\rm Edd}$, the corresponding power spectrum and the dynamical power spectrum with mass injection rate $\dot M_{\rm inject} = 2\dot M_{\rm Edd}$. 
    Left panel: $\dot M_{\rm in}/\dot M_{\rm Edd}$ as a function of $t$ at $r=2R_{\rm S}$ (top) and $r=3.8R_{\rm S}$ (bottom) respectively. The purple region and the red region refer to phase 1 ($20-30$) day and phase 2 ($33-47$) day respectively. Middle panel: Power spectrum calculated with FFT for the full data, phase I data and phase 2 data at $r=2R_{\rm S}$ (top) and $r=3.8R_{\rm S}$ (bottom) respectively. Right panel: Dynamical power spectrum calculated with FFT at $r=2R_{\rm S}$ (top) and $r=3.8R_{\rm S}$ (bottom) respectively.}
    \label{fig:6}
\end{figure*}

In the left panel of Fig. \ref{fig:6}, we plot the scaled mass inflow rates $\dot M_{\rm in}/\dot M_{\rm Edd}$ as a function of time $t$ at $r=2R_{\rm S}$ (top) and $r=3.8R_{\rm S}$ (bottom) respectively for mass injection rate $\dot M_{\rm inject} = 2\dot M_{\rm Edd}$. There are significant state transitions of $\dot M_{\rm in}/\dot M_{\rm Edd}$ between high states and low states after the accreted mass fills the gap between the mass injection radius and the BH. Since the behaviors of the state transitions are similar throughout the simulation, we select two parts of the data, i.e., phase 1 ($20-30$) day for high state and phase 2 ($33-47$) day for low state for $r=2R_{\rm S}$ as examples for the timing study. As for $r=3.8R_{\rm S}$, we also select $20-30$ day as phase 1 data for the high state although a few low state data has been included, and $33-47$ day as phase 2 data for low state. By performing FFT with phase 1 data ($20-30$) day, there is no QPO found for $r=2R_{\rm S}$, while there is a QPO frequency near $1.18\times 10^{-4}$ at $r=3.8R_{\rm S}$. Performing FFT with phase 2 data ($33-47$) day and full data ($0-400$) day, there are QPOs found at both $r=2R_{\rm S}$ and $r=3.8R_{\rm S}$. Specifically, the QPO frequency $\nu_{\dot M_{\rm in}}$ is $1.08\times 10^{-4}$ for phase 2 and full data at both $r=2R_{\rm S}$ and $r=3.8R_{\rm S}$. One can see the blue, red and the black symbols in the middle panel of Fig. \ref{fig:6} for details. The dynamical power spectrum is plotted in the upper right panel of Fig. \ref{fig:6} for $r=2R_{\rm S}$ and in the lower right panel of Fig. \ref{fig:6} for $r=3.8R_{\rm S}$ respectively. In the calculation, the time bin is also chosen to be 10 days. Comparing the dynamical power spectrum with $\dot M_{\rm in}/\dot M_{\rm Edd}$ in Fig. \ref{fig:6}, we notice that the QPO frequency only appears in the low state, while there is no QPO in the high state. The appearance of QPO at $r=3.8R_{\rm S}$ for the high state ($20-30$) day is because some low state date is included. 

The appearance of QPO at low state can be easily understood as that the matter in the accretion flow is very well constrained in the equatorial plane for $\dot M_{\rm inject}=2 \dot M_{\rm Edd}$ as for $\dot M_{\rm inject}=0.01 \dot M_{\rm Edd}$ and $\dot M_{\rm inject}=0.1 \dot M_{\rm Edd}$, in which the radial oscillation naturally can occur. However, if the accretion flow is in the high state, as shown in the second panel (the 20 day) of Fig. \ref{fig:3}, the accretion flow is extended in the vertical direction. In this case, the radial oscillation will be weaken. If the extension in vertical direction is significant, the radial oscillation can even disappear. The QPO frequencies for different radii are given in Table \ref{tab:3} and FFT parameters are given in the note.

\subsection{Theoretical analysis} \label{subsec:3.3}
In order to analyze the global QPO frequency of the accretion disk, we further plot the scaled mass inflow rates $\dot M_{\rm in}/\dot M_{\rm Edd}$ as a function of $t$, the power spectrum and the dynamical power spectrum at $r=2,3,3.5,3.8,5,10,15$ and $20R_{\rm S}$ for $M_{\rm inject} = 0.01 \dot M_{\rm Edd}$ and $\dot M_{\rm inject} = 2 \dot M_{\rm Edd}$ in Fig. \ref{fig:9} and Fig. \ref{fig:11} respectively, and at $r=2,3,3.5,3.8,5,10$ and $15R_{\rm S}$ for $\dot M_{\rm inject} = 0.1 \dot M_{\rm Edd}$ in Fig. \ref{fig:10}.
For $\dot M_{\rm inject} = 0.1 \dot M_{\rm Edd}$ we do not take $r=20R_{\rm S}$ because the main body of the accretion disk has not diffused to $20R_{\rm S}$ in the simulation, which can be seen in the right panel of Fig. \ref{fig:2}
for $t=300$ day as an example. Similar to Figs \ref{fig:4}, \ref{fig:5} and \ref{fig:6}, there are significant QPOs at these radius in the region of the formed accretion disks.

\begin{figure}
    \centering
    \includegraphics[width=0.45\textwidth]{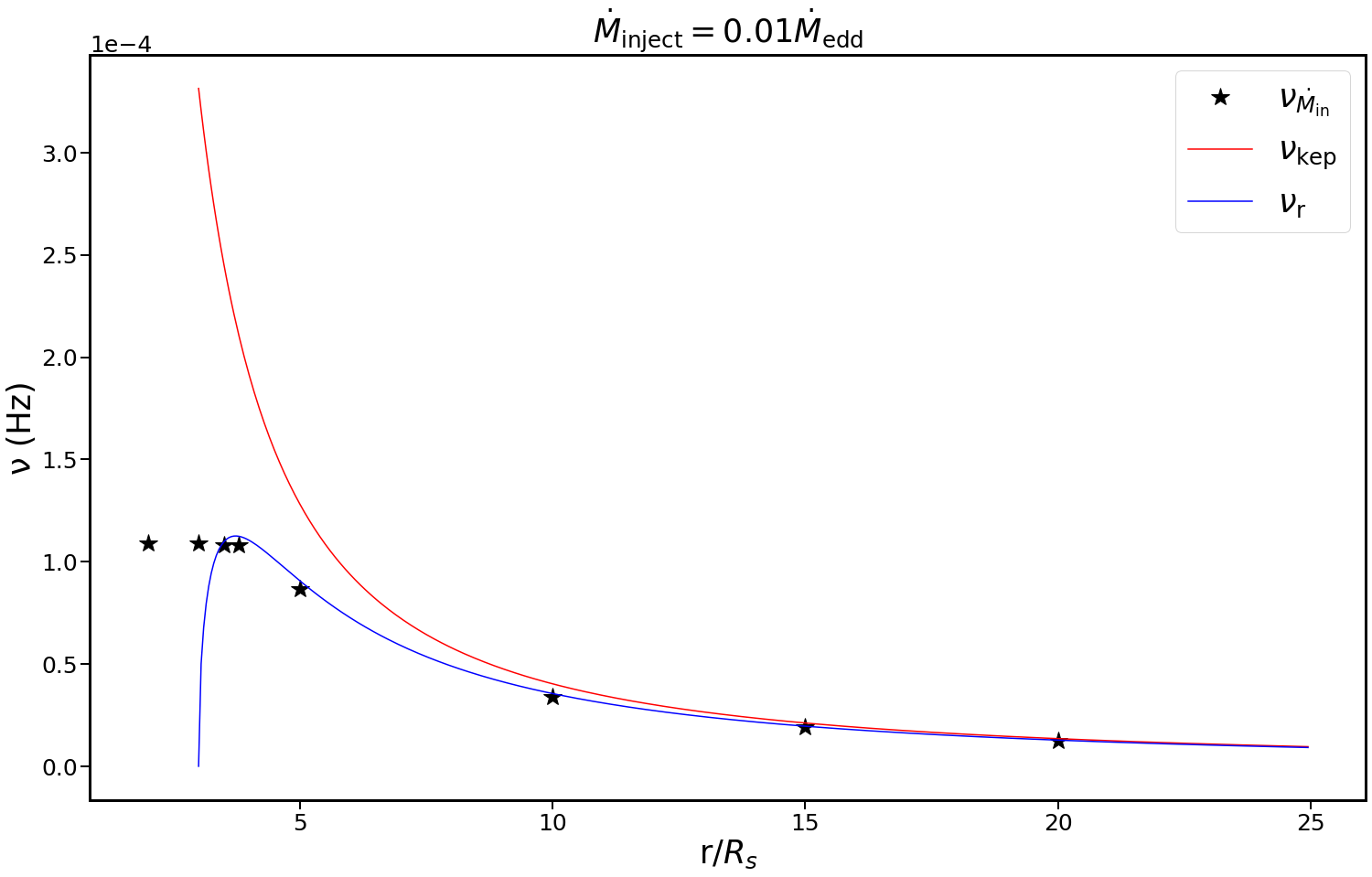}
    \hfill
    \includegraphics[width=0.45\textwidth]{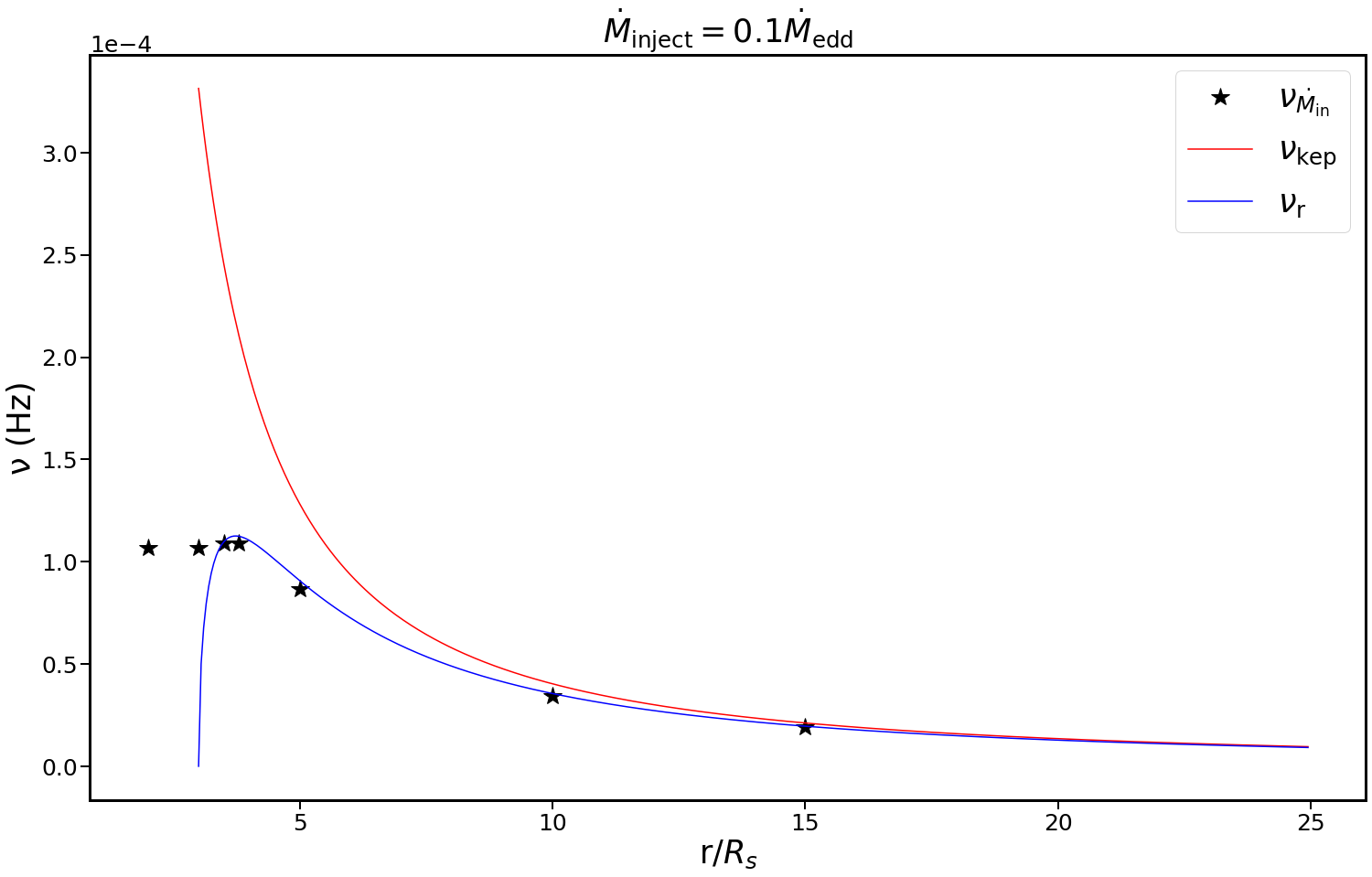}
    \hfill
    \includegraphics[width=0.45\textwidth]{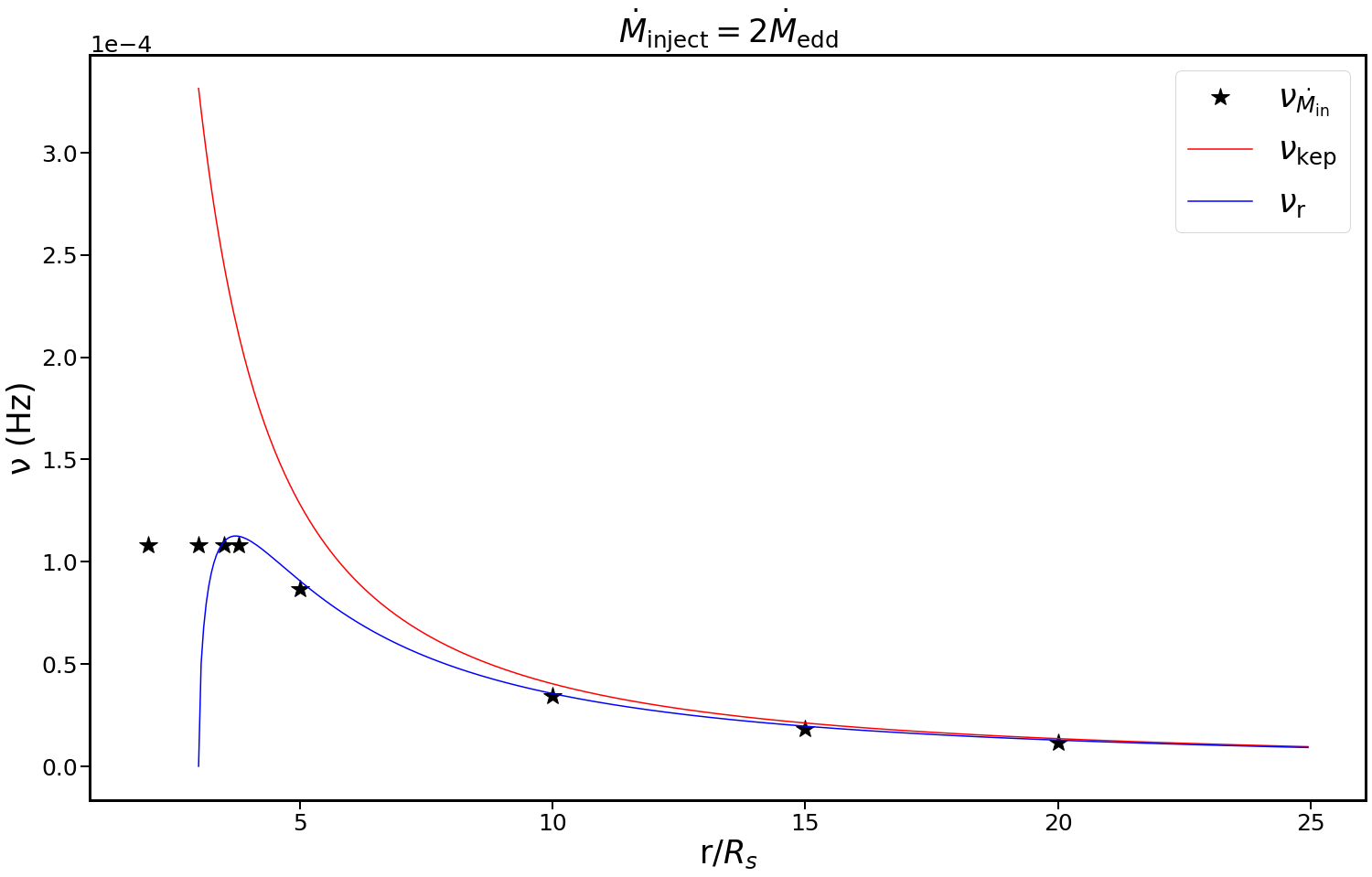}
    \caption{The theoretical Keplerian orbital frequency, the epicyclic frequency of the accretion disk, as well as the QPO frequency derived from our simulations as a function of radius $r/R_{\rm S}$ for $\dot M_{\rm inject} = 0.01 \dot M_{\rm Edd}$ (top), $\dot M_{\rm inject} = 0.1 \dot M_{\rm Edd}$ (middle) and $\dot M_{\rm inject} = 2 \dot M_{\rm Edd}$ (bottom) respectively. The red solid lines and the blue solid line represent the theoretical Keplerian orbital frequency (equation (\ref{eq:3})) and the epicyclic frequency of the accretion disk (equation (\ref{eq:4})) assuming the PN potential for a BH mass of $10^7 M_{\odot}$. The black stars represent the QPO frequency derived from our simulations of scaled mass inflow rates $\dot M_{\rm in}/\dot M_{\rm Edd}$ as a function of $t$ for $\dot M_{\rm inject} = 0.01 \dot M_{\rm Edd}$ (top), $\dot M_{\rm inject} = 0.1 \dot M_{\rm Edd}$ (middle) and $\dot M_{\rm inject} = 2 \dot M_{\rm Edd}$ (bottom) respectively.}
    \label{fig:7}
\end{figure}

We plot the Keplerian orbital frequency  $\nu_{\rm kep}$, the epicyclic frequency of the accretion disk $\nu_{\rm r}$, as well as the QPO frequency derived from our simulations $\nu_{\dot M_{\rm in}}$ as a function of radius $r/R_{\rm S}$ in Fig. \ref{fig:7}, where the red solid lines represent the theoretical Keplerian orbital frequencies in the PN potential for a BH with a mass of $10^7M_{\odot}$, the blue solid lines represent the theoretical radial epicyclic frequencies of the accretion disk in the PN potential from analytic calculations for a BH with a mass of $10^7M_{\odot}$, and the black stars represent the QPO frequency derived from our simulations for $\dot M_{\rm inject} = 0.01 \dot M_{\rm Edd}$ (top), $\dot M_{\rm inject} = 0.1 \dot M_{\rm Edd}$ (middle) and $\dot M_{\rm inject} = 2 \dot M_{\rm Edd}$ (bottom) respectively.
Specifically, $\nu_{\rm kep}$ and $\nu_{\rm r}$ can be expressed as follows,

\makeatletter
\@fleqnfalse
\makeatother
\begin{equation}
\nu_{\rm kep}=\frac{1}{2\pi}*\sqrt{\frac{GM_{\rm BH}}{r}}*\frac{1}{r-R_{\rm s}}
\label{eq:3}
\end{equation}

\begin{equation}
\nu_{\rm r}=\frac{1}{2\pi}*\sqrt{\frac{GM_{\rm BH}}{r}}*\frac{1}{r-R_{\rm s}}*\sqrt{\frac{r-3R_{\rm s}}{r-R_{\rm s}}}.
\label{eq:4}
\end{equation}
\makeatletter
\@fleqntrue
\makeatother

We notice that the three panels in Fig. \ref{fig:7} for different $\dot M_{\rm inject}$ in the simulations share very similar signatures. 
The derived QPO frequencies at different radii from our simulations are well consistent with the epicyclic frequency (blue solid line) predicted by equation (\ref{eq:4}) at radius greater than $3.8R_{\rm S}$, where the radial epicyclic frequency reaches its maximum value. While the derived QPO frequencies from the simulation nearly keep constant with decreasing radius from $r=3.8R_{\rm S}$ to $r=2R_{\rm S}$. We also found that although the derived QPO frequencies in the region less than $3.8R_{\rm S}$ nearly keep constant with decreasing radius, the power of the QPO frequencies decreases significantly with decreasing radius. One can see Fig. \ref{fig:9}, \ref{fig:10} and \ref{fig:11} for details. 

By derivating to equation (\ref{eq:4}), it is easy to find that the radius where the maximum frequency of equation (\ref{eq:4}) is at $r=(2+\sqrt{3})R_{\rm S}\approx 3.8R_{\rm S}$. We think that the nearly constant QPO frequency and
the significant decreased power of the QPO can be simply understood as that 
the maximum radial epicyclic frequency at $3.8R_{\rm S}$ is being propagated inward to the inner boundary of the accretion disk, and partly dissipated in the course of the propagation. Since the oscillation of the mass inflow rates in the innermost region of the accretion disk can naturally reflect the observed QPO
frequency $\nu_{\rm QPO}$, we suggest that the theoretical maximum radial epicyclic frequency, i.e  the radial epicyclic frequency at $3.8R_{\rm s}$ is a good proxy for indicating $\nu_{\rm QPO}$ as also suggested by \citet[][]{Kluzniak&Abramowicz2001}.
Specifically, according to equation (\ref{eq:4}), this maximum radial epicyclic frequency can be expressed as,

\makeatletter
\@fleqnfalse
\makeatother
\begin{equation}
\nu_{\rm r,max}=\nu_{\rm r}((2+\sqrt3) R_{\rm s})=\frac{1}{2\pi}*\sqrt{\frac{GM_{\rm BH}}{(2+\sqrt3) R_{\rm s}}}*\frac{1}{(1+\sqrt3) R_{\rm s}}*\sqrt{\frac{\sqrt3-1}{\sqrt3+1}}.
\label{eq:5}
\end{equation}
\makeatletter
\@fleqntrue
\makeatother

\begin{figure}
    \centering
    \includegraphics[width=0.45\textwidth]{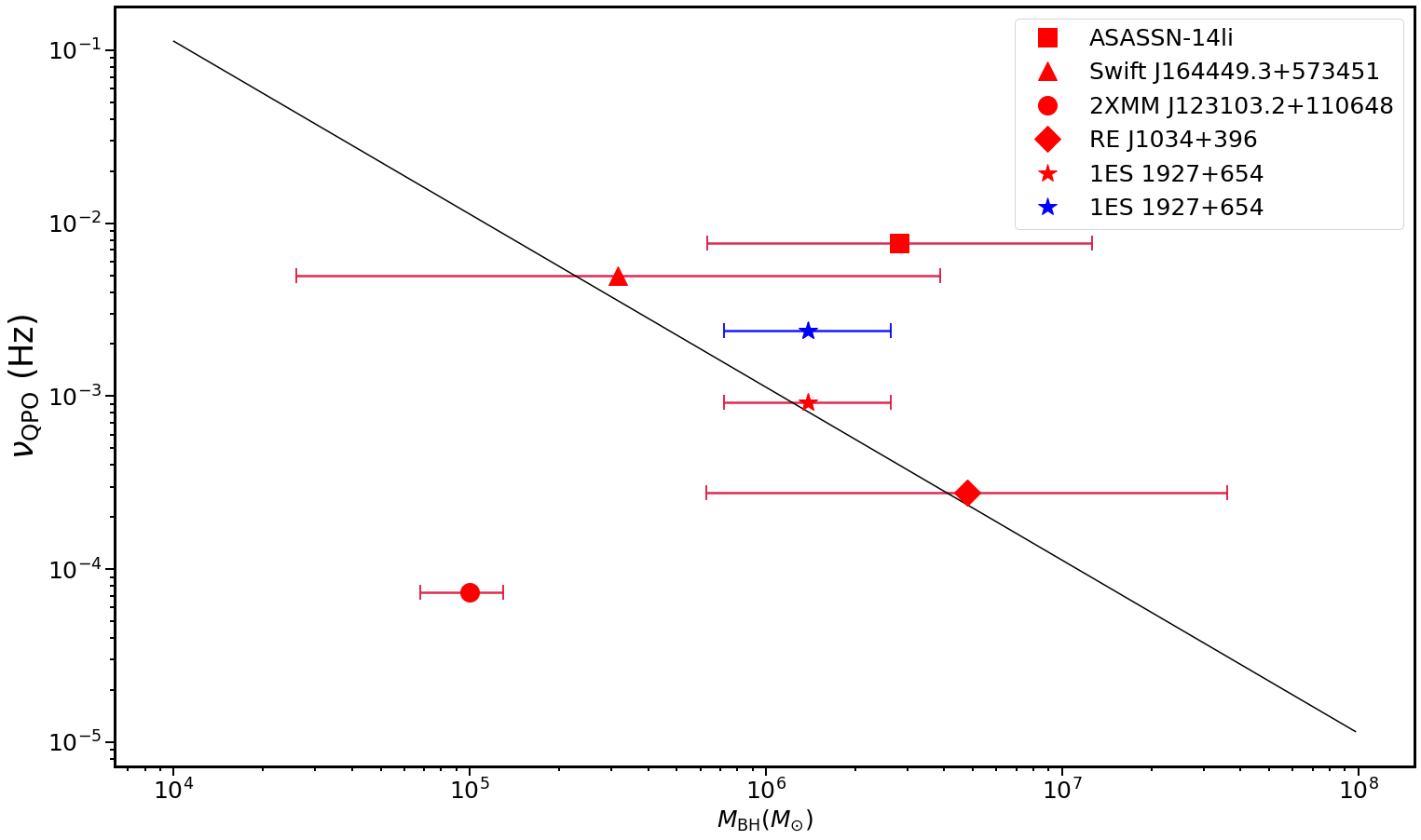}
    \caption{Relation between the observed QPO frequency $\nu_{\rm QPO}$ and the BH mass $M_{\rm BH}$ for five sources, including two AGNs, i.e, RE J1034+396, 1ES 1927+654, as well as three TDEs, i.e., Swift J164449.3+573451, ASASSN-14li and 2XMM J123103.2+110648 (TDE candidate). The black solid line is the maximum radial epicyclic frequency $\nu_{\rm r, max}$ as a function of $M_{\rm BH}$, i.e equation (\ref{eq:5}) in the paper. }
    \label{fig:8}
\end{figure}

We search for SMBH systems with QPO detections from the literatures. For SMBH systems, we consider both AGN and TDE. In Fig. \ref{fig:8}, we plot $\nu_{\rm QPO}$ versus the measured BH mass for five sources, including two AGNs, i.e, RE J1034+396 \citep{Gierlinski2008}, 1ES 1927+654 \citep[][]{Masterson2025,Li2022}\footnote{For 1ES 1927+654, an evolution of the QPO frequency has been observed. Specifically, two QPO frequencies have been detected, which are plotted in Fig. \ref{fig:8} with red star and blue star respectively.}, as well as three TDEs, i.e., Swift J164449.3+573451 \citep[][]{Reis2012,Miller&Gultekin2011}, ASASSN-14li \citep{Pasham2019} and 2XMM J123103.2+110648 (TDE candidate) \citep{Lin2013}. We also plot the $\nu_{\rm r,max}$ as a function of $M_{\rm BH}$, i.e., equation (\ref{eq:5}) for comparison. It can be seen that except for 2XMM J123103.2+110648 and ASASSN-14li the observed QPO frequency $\nu_{\rm QPO}$ as a function of $M_{\rm BH}$ is roughly consistent with equation (\ref{eq:5}), which means that maximum radial epicyclic frequency of the accretion disk can indeed indicate the observed QPO frequency for an accreting SMBH.

The observed QPOs in 2XMM J123103.2+110648 and ASASSN-14li may have different physical mechanisms compared with that of others. In \citet{Zhou2015}, the authors collected a sample of BH accreting systems, including black hole X-ray binaries (GRO J1655-40, XTE J1550-64, GRS 1915+105, and H 1743-322), ultra-luminous X-ray source (M82 X-1, NGC 5408 X-1), Seyfert galaxy (RE J1034+396), TDE (Swift J164449.3+573451) and Sgr $\rm A^{*}$ with QPO detections, in which the detected QPO for black hole X-ray binaries are classified as HFQPO. Further, the author plotted the observed relation between $\nu_{\rm QPO}$ and $M_{\rm BH}$, it is claimed that there is a universal scaling relation between $\nu_{\rm QPO}$ and $M_{\rm BH}$. Consequently, it is suggested that the observed QPO in M82 X-1, NGC 5408 X-1, RE J1034+396, Swift J164449.3+573451 and Sgr $\rm A^{*}$ are also HFQPO. As for 2XMM J123103.2+110648, the observed QPO frequency is much lower than the prediction of the suggested scaling relation, which probably has a different physical origin such as thermal instability \citep{Lin2013}.
For ASASSN14-li, the observed QPO frequency is higher than the prediction of the suggested scaling relation, which suggests that the central BH could have a moderate spin. In this case, the radiating material producing the QPO is located closer to event horizon of the BH. 
However, even if the effect of the BH spin to the maximum radial epicyclic frequency is considered (see Section \ref{subsec:4.1}), the observed QPO frequency of ASASSN14-li is still higher than the theoretical prediction, as shown in 
Fig. \ref{fig:12}, which implies that this QPO may represent a different disk oscillation mode from other systems \citep{Pasham2019}.

We should notice that in the present paper, we perform the simulations by fixing BH mass at $10^{7}M_{\odot}$. 
It is not very clear whether the predicted QPO frequencies as a function of radius from the simulations still can well match the radial epicyclic frequency as a function of radius from analytic calculations for different BH masses. If the simulation results can be scaled to different BH masses, equation (\ref{eq:5}) can be used to compare with observations.
In general, the geometry and the dynamics of the accretion flow around a BH is determined by the scaled mass accretion rate $\dot m$ (defined as $\dot m=\dot M/\dot M_{\rm Edd}$) both in black hole X-ray binaries \citep[][]{Esin1997,Qiao2013} and AGNs \citep[][]{Ho2008ARA&A}, while the effects of BH mass to the the geometry and the dynamics of the accretion flow are relatively minor \citep[][]{Merloni2003,Falcke2004}. 
Meanwhile, in the present paper, the QPO sources in our sample spans a relative narrow range of BH mass ($\sim 10^{5-7}M_{\odot}$), between which the behaviors 
of the accretion physics are expected to not change much for a fixed $\dot m$. Finally, as for the precise calculations of the effects of BH mass to the properties of the accretion flow, we expect to perform simulations in the future work.

\section{Discussion} \label{sec:4}
\subsection{The effect of different gravitational field forms} \label{subsec:4.1}

In this paper, we perform radiation hydrodynamic simulation with pseudo-Newtonian potential to mimic the effect of the general relativity around a Schwarzschild BH.
We find that the derived QPO frequencies of the mass inflow rates at different radii from our simulations are well consistent with the radial epicyclic frequency predicted by analytic calculations for radius greater than 3.8 $R_{\rm S}$, where the radial epicyclic frequency reaches its maximum value.
We then propose that the maximum radial epicyclic frequency of the mass inflow rate can be a good proxy for the observed QPO frequency as discussed in Section \ref{subsec:3.3}. Further, if we assume that the maximum radial epicyclic frequency from analytic calculation in general relativity case can also indicate the observed QPO frequency and can be scaled for different BH masses, we can plot the maximum radial epicyclic frequency based on the analytic formula as a function of BH mass as a comparison. In the case of general relativity, the radial epicyclic frequency was given by \citet{Nowak1998},

\makeatletter
\@fleqnfalse
\makeatother
\begin{equation}
\nu_{\rm r} = \sqrt{\alpha_{\rm r}}\nu_{\rm kep}
\label{eq:6}
\end{equation}

\begin{equation}
\nu_{\rm kep} = \frac{1}{2\pi}\sqrt{\frac{GM_{\rm BH}}{{r_{\rm g}}^3}}{(x^{\frac{3}{2}}+a)}^{-1}
\label{eq:7}
\end{equation}
\makeatletter
\@fleqntrue
\makeatother

\noindent where $\alpha_{\rm r} = 1-6x^{-1}+8ax^{-\frac{3}{2}}-3a^2x^{-2}$, $x=\frac{r}{r_{\rm g}}$, $r_{\rm g}=\frac{GM_{\rm BH}}{c^2}$, $a=\frac{J_{\rm BH}c}{G{M_{\rm BH}}^2}$, $J_{\rm BH}$ represents the internal angular momentum of the central BH, $a$ is the spin parameter. 
Substituting equation (\ref{eq:7}) into equation (\ref{eq:6}), we numerically search for $\nu_{\rm r, max}$ for a fixed $M_{\rm BH}$. 
We then plot $\nu_{\rm r, max}$ as a function of $M_{\rm BH}$ for $a=0$ and $a=0.998$, which can been seen in Figure \ref {fig:12} for the dotted line and the dashed line respectively. 
It can be seen that for a fixed $\nu_{\rm r, max}$, the derived $M_{\rm BH}$ from PN potential is roughly in the middle of the cases taking $a=0$ and $a=0.998$ from GR respectively. The BH mass derived from PN potential is roughly 1.5 times larger than that of $a=0$ from GR, and 2 times lower than that of $a=0.998$ from GR. Considering the error for BH mass measurement, we think that PN potential is a good approximation for the purpose in the present simulations. Even so, we think it is very necessary to introduce GR in simulations to more accurately reflect the GR effect, which will be done in the future.

Furthermore, if we consider a misaligned disk around the rotating central black hole, which is believed to be the most common state in real TDEs, the tilted disk will experience a Lense-Thirring torque from the BH which may dominate over the dynamics, and may be teared and result in a more complicated evolution \citep{Liska2021,Musoke2023}. In \citet{Musoke2023}, the author simulated the evolution of a thin accretion disk around a rapidly spinning BH with $a=0.9375$. In the simulation, they tilted the disk by $65^{\circ}$ with respect to the horizontal equatorial plane of the BH under their initial conditions. From this simulation, they found a low precession frequency from the teared inner sub-disks caused by Lense-Thirring torque from the central BH, while the radial epicyclic oscillations still exist in the inner disk. That is to say, while the Lense-Thirring torque dominate the large scale motion of the disk, the epicyclic oscillations discussed in this paper will still show HFQPO-like frequencies.

Compared with BH mass, the constraints to spin for the SMBH sources mentioned in this paper is much more uncertain. 
For ASASSN-14li, it is suggested to have a high spin, since the detected $\sim7.65$ mHz QPO signal requires emission originating from a region very close to the innermost stable circular orbit of a rapidly spinning BH if the independently estimated BH mass is adopted \citep{Pasham2019}.
For RE J1034+396, spectral studies suggest a high accretion rate, but current X-ray spectral constraints do not uniquely determine the spin \citep{Gierlinski2008,Alston2014}.
For 1ES 1927+654 and 2XMM J123103.2+110648, spin estimates are also weak. It is suggested that QPO frequencies can be used to constrain the spin, which however remain uncertainties \citep{Masterson2025,Lin2013,lin2017}.
For Swift J164449.3+573451, the presence of a relativistic jet is often  interpreted as a direct evidence for a rapidly rotating BH, but the precise measurement of spin is still lack \citep{Burrows2011,Bloom2011,Levan2011,Reis2012}. 
In this paper, we propose that the observed QPO frequency can be interpreted by the maximum radial epicyclic frequency of the accretion flow. In this framework, the BH spin can in principle be constrained once the BH mass is independently estimated.

\subsection{Other possible factors affecting the QPO signal}\label{subsec:4.2}

\subsubsection{The effect of mass injecting radius}\label{subsubsec:4.2.1}
In this paper, we fix the mass injecting radius at $r=10R_{\rm S}$ in the simulations as we do not expect the viscous timescale to be too long for the accretion flow to diffuse to the BH.
For such a smaller injecting radius, it allows the accretion flow to reach a quasi-steady state rapidly, which is necessary for examining the development of the oscillation near the BH.
However, in the actual AGN environment the accretion disk has a much larger spatial scale in radial direction, which is about several thousands $R_{\rm S}$. 
If we put our mass injecting point to $\sim 1000R_{\rm S}$, we can expect a much longer viscous timescale for the injected matter to evolve to the inner boundary of the BH, i.e., $2R_{\rm S}$, which would be a strong challenge to the hardware we use for our simulations. In this case, the formation time of the inner accretion flow itself would be prolonged, which may also delay or weaken the appearance of a stable QPO signal in the inner region.
Despite the difficulty of setting a larger mass injecting radius in the simulation, it is still necessary to test the effect of mass injecting radius to the simulation results, which are expected to be done in the future. 

One of the possible applications for setting different mass injecting radius is to explain the observed QPO detection rates between AGNs and TDEs. As we know, the QPO detection rate in AGNs are much lower than that in TDEs. Although the physical reasons of how a larger mass injecting radius can affect the production of the QPO is unclear, we can simply analyze the reasons as follows. As has been discussed in Section \ref {subsec:3.3}, the QPOs detected in AGNs and TDEs are suggested to be HFQPOs, which are believed to be occurred in the innermost region around a SMBH. If the mass injecting radius is enlarged, it is very possible that the strength of the epicyclic motion in the innermost region is diluted considering the global motion of the accretion disk. In the meantime, the observed QPO signal should be calculated by integrating the whole accretion disk, an increase of the size of the accretion disk will make the relative contribution of the inner disk to the global disk decreased, consequently make the strength of the QPO signals decreased or even disappeared.

\subsubsection{The effect of viscosity}\label{subsubsec:4.2.2}

The viscosity parameter may also influence the QPO signal. Some semi-analytical studies suggest that stronger viscosity can enhance radial oscillatory motions in the inner region of the accretion disk through viscous overstability or related viscous instabilities under certain conditions \citep{Kato1978,Honma1992,Chen1995}, although the characteristic frequency is still expected to remain close to the local radial epicyclic frequency.
In such a case, viscosity is more likely to modify the shape of the peak in the power spectrum, rather than significantly shifting the peak frequency itself.
In the contrast, using 2D HD simulations, \citet{Miranda2015} suggests that for overstable trapped p modes, the effect of viscosity is either to reduce the growth rates of these modes, or to completely suppress them and excite a new class of higher frequency modes. The latter is a result of the classic viscous overstability effect. 
Therefore, exploring QPO signals under different viscosity parameters $\alpha$ would be valuable for establishing a possible connection between the power spectrum properties of QPOs and the viscous properties of the accretion flow.

\subsubsection{The effect of magnetic field}\label{subsubsec:4.2.3}

In the present paper, we neglect magnetic field in our simulations, which may affect the properties of the QPO. 
In \citet{Neill2009}, the author compared the viscous HD simulation and the MHD simulation, finding that trapped global mode oscillations in the $\alpha$-disk are not easily identified in the MHD disk simulations. 
Nevertheless, recent GRMHD simulation of a tilt thin disk suggest that radial epicyclic oscillation mode may still survive to explain the observed HFQPO in BHXBs \citep{Musoke2023}.
In our simulation, for $\dot M_{\rm inject} = 0.01 \dot M_{\rm Edd}$ and $\dot M_{\rm inject} = 0.1 \dot M_{\rm Edd}$ the accretion flow fall in the thin disk regime. In these cases, if the magnetic field is not included, the wind is weak as shown in Fig. \ref{fig:1} and Fig. \ref{fig:2}. If a large-scale poloidal magnetic field is considered, a magneto-centrifugally driven wind will be expected by the Blandford-Payne effect \citep{Blandford1982,Lesur2013,Vourellis2019,Dihingia2021,Liska2022,Dihingia2024}. If there exists strong wind, the QPO derived from $\dot M_{\rm in}$ as a function of $t$ will be smeared to some extent as discussed in Section \ref{subsec:4.3}.
For $\dot M_{\rm inject} = 2 \dot M_{\rm Edd}$, the accretion flow fall in the slim disk regime, where the wind is driven by the radiation pressure as show in Fig. \ref{fig:3}. We should note that several studies have been carried out for the super-Eddington accretion flow with radiation MHD (RMHD) simulations. In these simulations, the wind is also dominated by radiation pressure \citep{Jiang2014,Dai2018,Jiang2019,Jiang2024}. Since RMHD simulation can more intrinsically reflect the properties of the accretion flow, it is necessary to perform RMHD simulations for the QPO properties for the super-Eddington accretion flows around SMBHs in the future, which however exceed the scope in the present paper.

\subsection{From mass inflow rate $\dot M_{\rm in}$ to the observables} \label{subsec:4.3}

In this paper, we show the QPO signals by analyzing the mass inflow rates $\dot M_{\rm in}/\dot M_{\rm Edd}$ based on radiation hydrodynamic simulations of accretion flow around a SMBH. 
Although it can be expected that the variation of $M_{\rm in}/\dot M_{\rm Edd}$ can naturally reflect the observed variability and the QPO, we think the global calculation for the emergent spectrum and the true light curve is still needed in the future. This is because the QPO signals calculated from $M_{\rm in}/\dot M_{\rm Edd}$ can be affected by other factors, such as outflows, which consequently will lead to some uncertainties when being compared with observations. In general, when the mass injection rate increases, the outflow is believed to increase in the meantime as has been shown in Section \ref {subsec:3.1} for $\dot M_{\rm inject} = 0.01 \dot M_{\rm Edd}$, $\dot M_{\rm inject} = 0.1 \dot M_{\rm Edd}$ and $\dot M_{\rm inject} = 2 \dot M_{\rm Edd}$ respectively.
Taking $\dot M_{\rm inject} = 2 \dot M_{\rm Edd}$ as an example, we can see at some time when the outflow is strong, the main part of the disk will be obscured by the outflows. In this case, the QPO signal will be diluted, consequently making the strength of the QPO decreased. 
As we can see from Section \ref {subsec:3.1}, the geometry of the outflow is non-spherically symmetric, i.e., along the line-of-sight for a smaller viewing angle, the outflow is optically thin and very weak, while for a larger viewing angle, the outflow gradually becomes optically thick and strong. 
So we can expect that if we observe the BH accreting system with a larger viewing angle, the QPO signature will be severely changed, and even is suppressed to be disappeared. 

In addition, the viewing angle is poorly constrained for most SMBH QPO sources mentioned in this paper. RE J1034+396, a narrow-line Seyfert 1 galaxy, is commonly considered to have a relatively low inclination. ASASSN-14li can be interpreted as being observed at low inclination because of its soft X-ray disk-like emission properties and short delay time between X-ray and optical/UV peaks \citep{Brown2017,Pasham2019}. 
Swift J164449.3+573451 is a jetted TDE, so it is believed that the viewing angle is very small assuming that the jet axis is perpendicular to the equatorial plane of the accretion disk \citep{Burrows2011,Bloom2011,Levan2011,Reis2012}. 
The viewing angle of 1ES 1927+654 and 2XMM J123103.2+110648 remain uncertain \citep{Masterson2025,lin2017}. 

In summary, the detailed calculation for the global emergent spectra and the corresponding multi-band light curves for different viewing angles, as well as the comparison with the specific source is very necessary in the future, which however exceeds the scope in the present paper.

\section*{Acknowledgements}
Yiyang Lin thanks the very useful discussions with Chenlei Guo and Wei Chen from NAOC on writing the manuscript. 
We thank Hongjun Zhong for technical support. 
We thank the participants of the TDE FORUM (Full-process Orbital to Radiative Unified Modeling) online seminar series for their inspiring discussions. 
This work is supported by the Strategic Priority Research Program of the Chinese Academy of Science (Grant No.XDB0550200), Shandong Provincial Key Research and Development Program (No.2022CXGC020106), National Super-computing Center in Jinan for computing resource, National Key R\&D Program of China (No.2023YFA1607903) and National Natural Science Foundation of China (Grant No. 12173048, 12333004).


\section*{Data availability}
The data underlying this article will be shared on reasonable request to the corresponding author.



\bibliographystyle{mnras}
\bibliography{example} 
\newpage




\appendix

\numberwithin{figure}{section}
\section{FFT results for $\dot M_{\rm in}$ at different radius} \label{sec:Appendix A}

\begin{figure*}
    \centering
    \includegraphics[width=0.49\textwidth]{figures/1M7-0.01Mdot/2Rs.png}
    \hspace{0.001\textwidth}
    \includegraphics[width=0.49\textwidth]{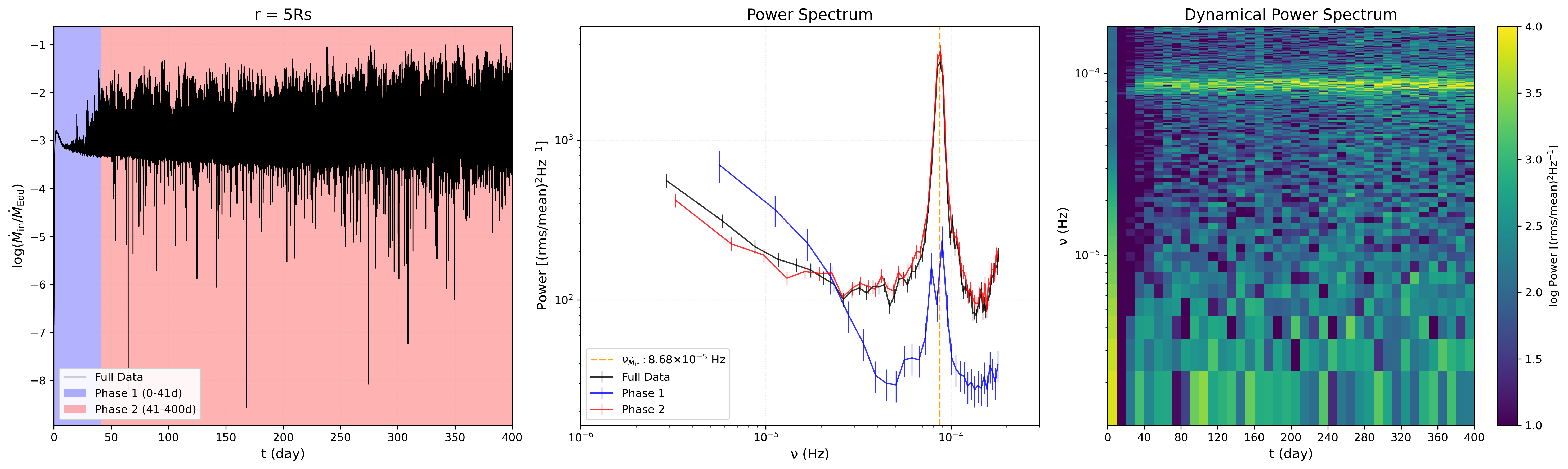}
    \hfill
    \includegraphics[width=0.49\textwidth]{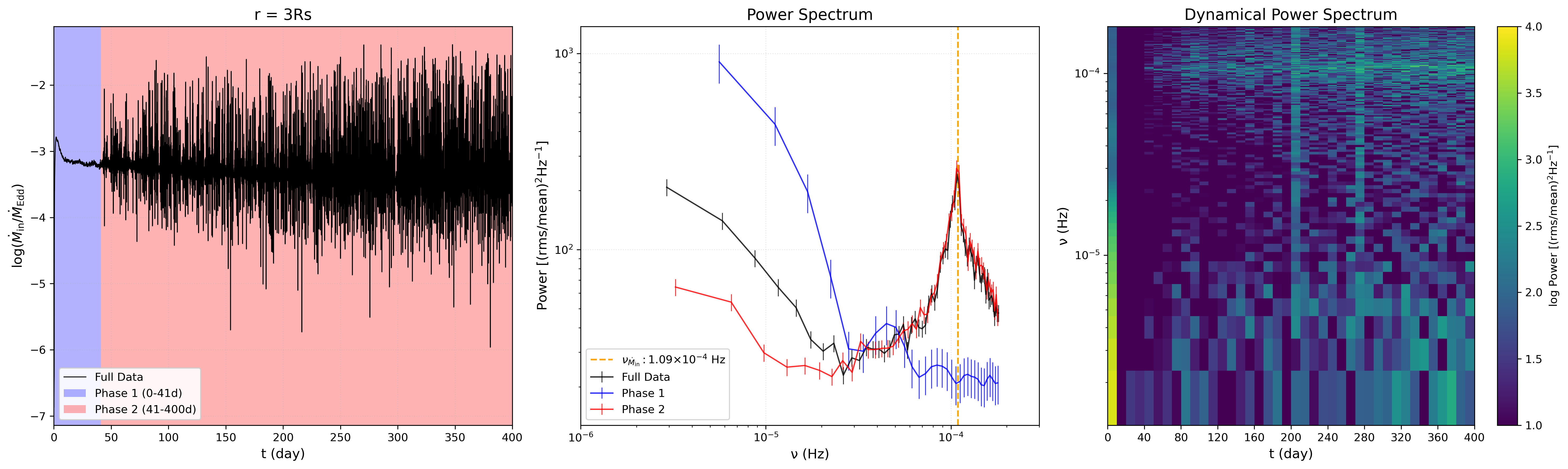}
    \hspace{0.001\textwidth}
    \includegraphics[width=0.49\textwidth]{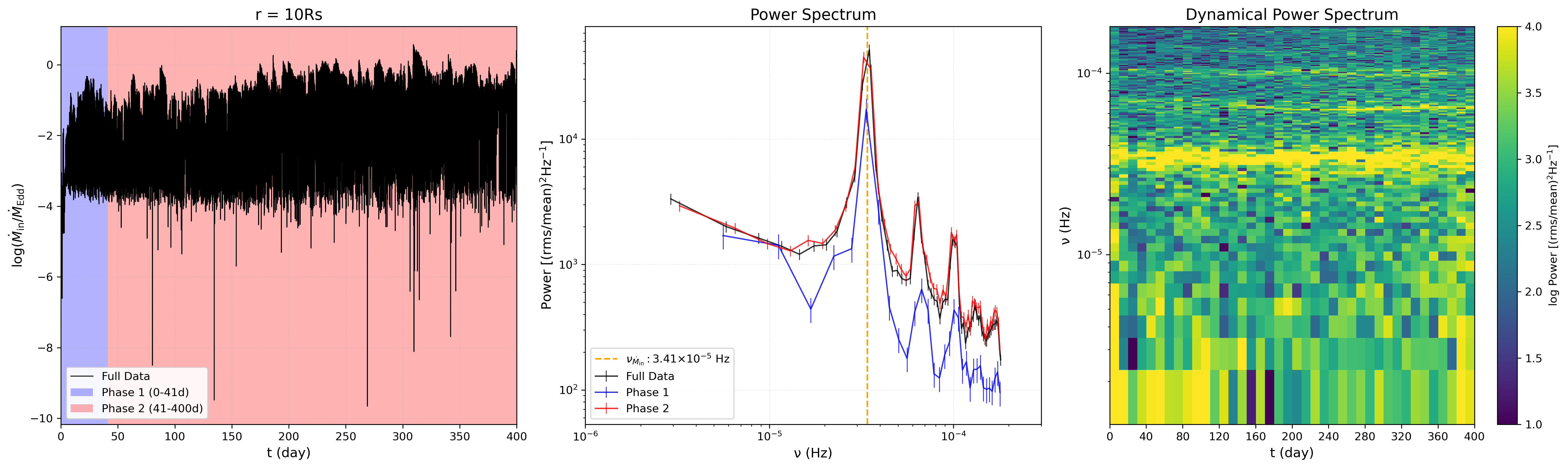}
    \hfill
    \includegraphics[width=0.49\textwidth]{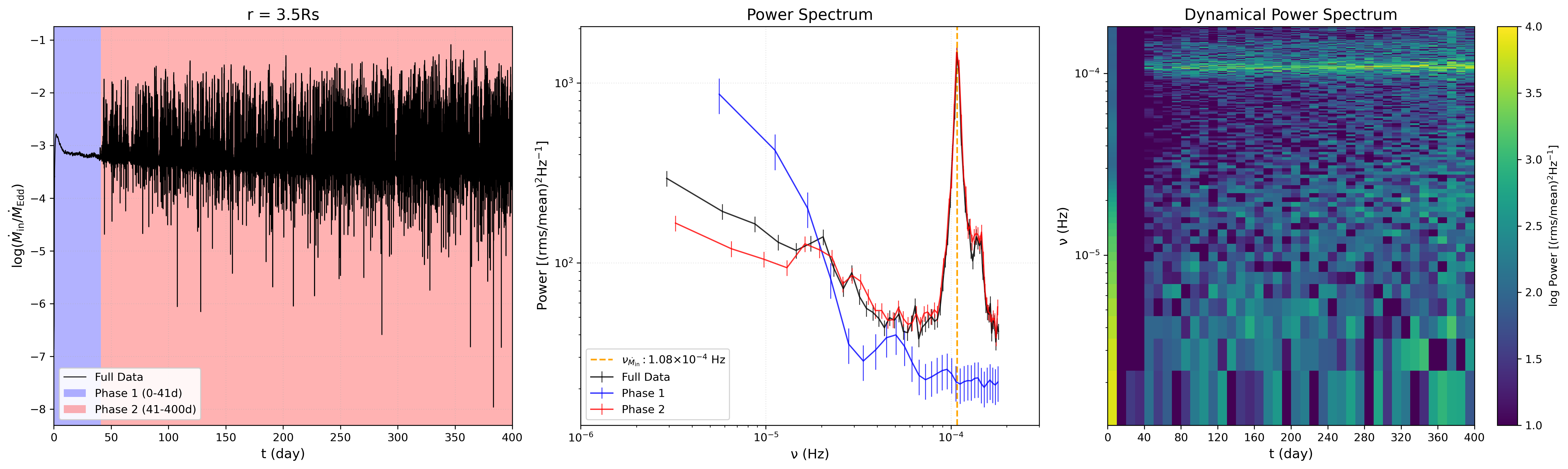}
    \hspace{0.001\textwidth}
    \includegraphics[width=0.49\textwidth]{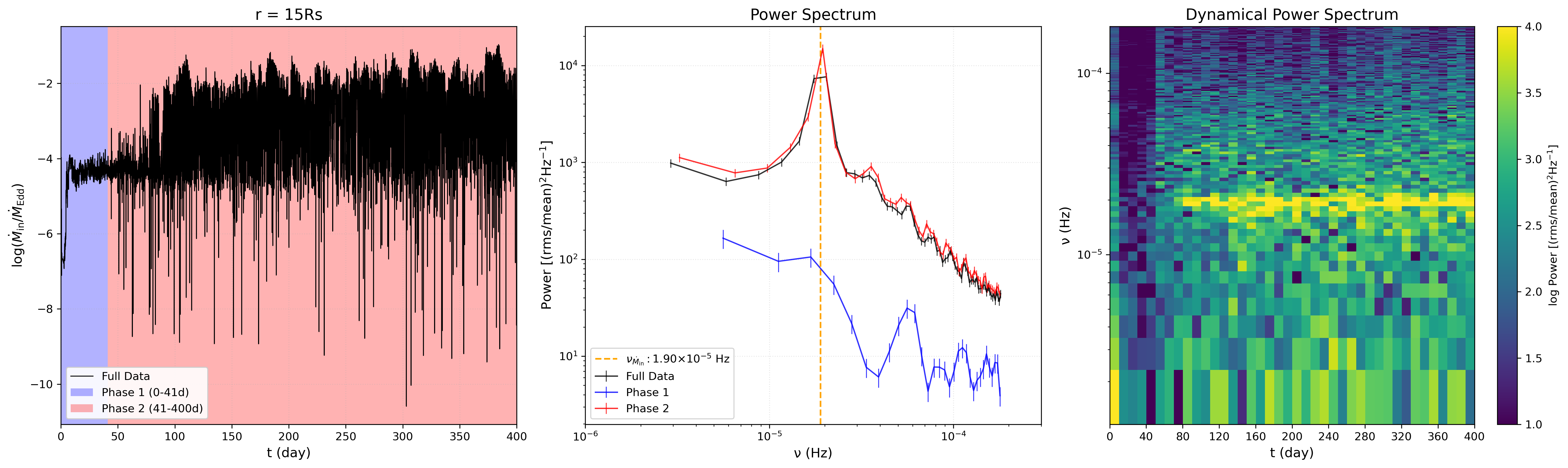}
    \hfill
    \includegraphics[width=0.49\textwidth]{figures/1M7-0.01Mdot/3.8Rs.png}
    \hspace{0.001\textwidth}
    \includegraphics[width=0.49\textwidth]{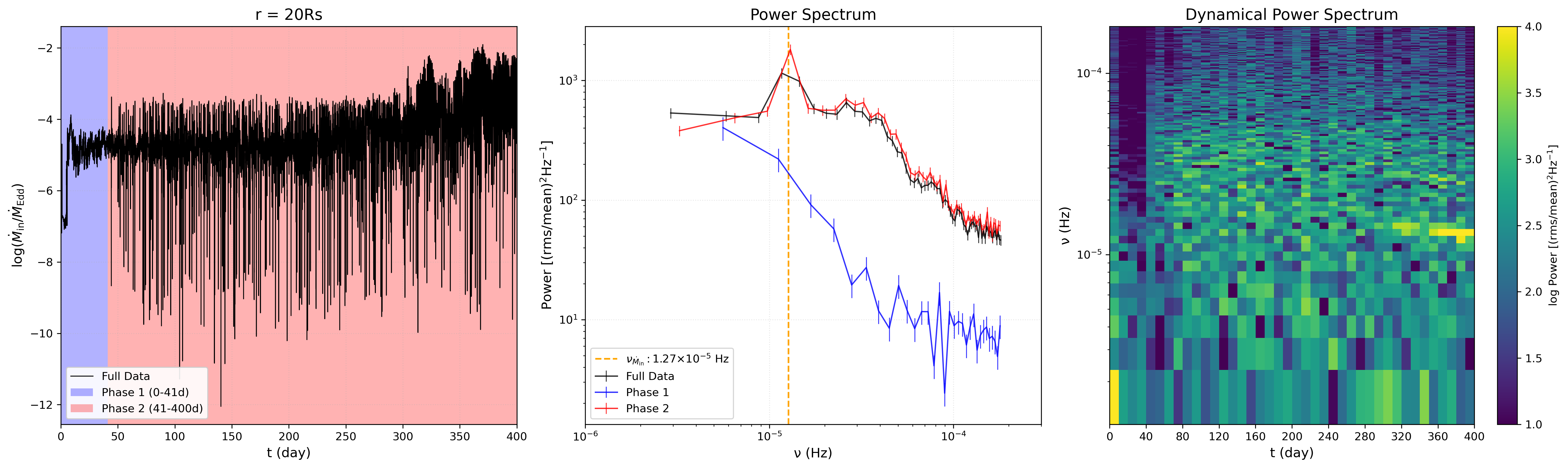}
    \caption{Evolution of the scaled mass inflow rate $\dot M_{\rm in}/\dot M_{\rm Edd}$, the corresponding power spectrum and the dynamical power spectrum with mass injection rate $\dot M_{\rm inject} = 0.01\dot M_{\rm Edd}$. Left side: results for $r=2R_{\rm S}$, $r=3R_{\rm S}$, $r=3.5R_{\rm S}$ and $r=3.8R_{\rm S}$ (from top to bottom). Right side: results for $r=5R_{\rm S}$, $r=10R_{\rm S}$, $r=15R_{\rm S}$ and $r=20R_{\rm S}$ (from top to bottom). For each figure at specific $r$, 
    left panel shows $\dot M_{\rm in}/\dot M_{\rm Edd}$ as a function of $t$  respectively. The purple region and the red region refer to phase 1 ($0-41$) day and phase 2 ($41-400$) day respectively; middle panel shows power spectrum calculated with FFT for the full data, phase I data and phase 2 data  respectively; right panel shows dynamical power spectrum calculated with FFT  respectively.}
    \label{fig:9}
\end{figure*}

\begin{figure*}
    \centering
    \includegraphics[width=0.49\textwidth]{figures/1M7-0.1Mdot/2Rs.png}
    \hspace{0.001\textwidth}
    \includegraphics[width=0.49\textwidth]{figures/1M7-0.1Mdot/3.8Rs.png}
    \hfill
    \includegraphics[width=0.49\textwidth]{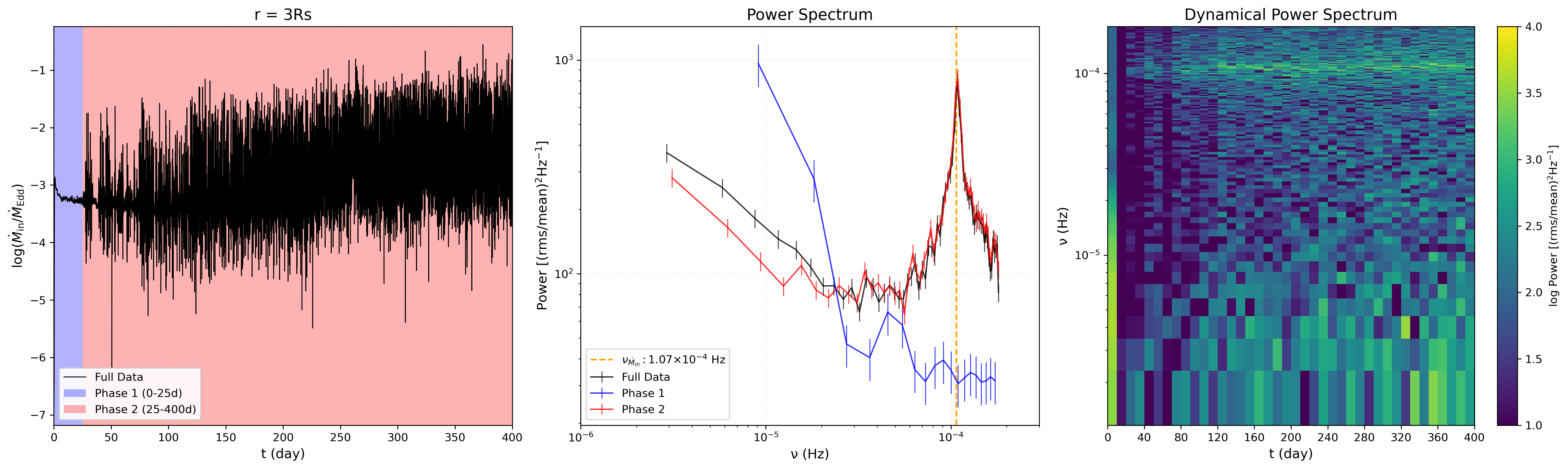}
    \hspace{0.001\textwidth}
    \includegraphics[width=0.49\textwidth]{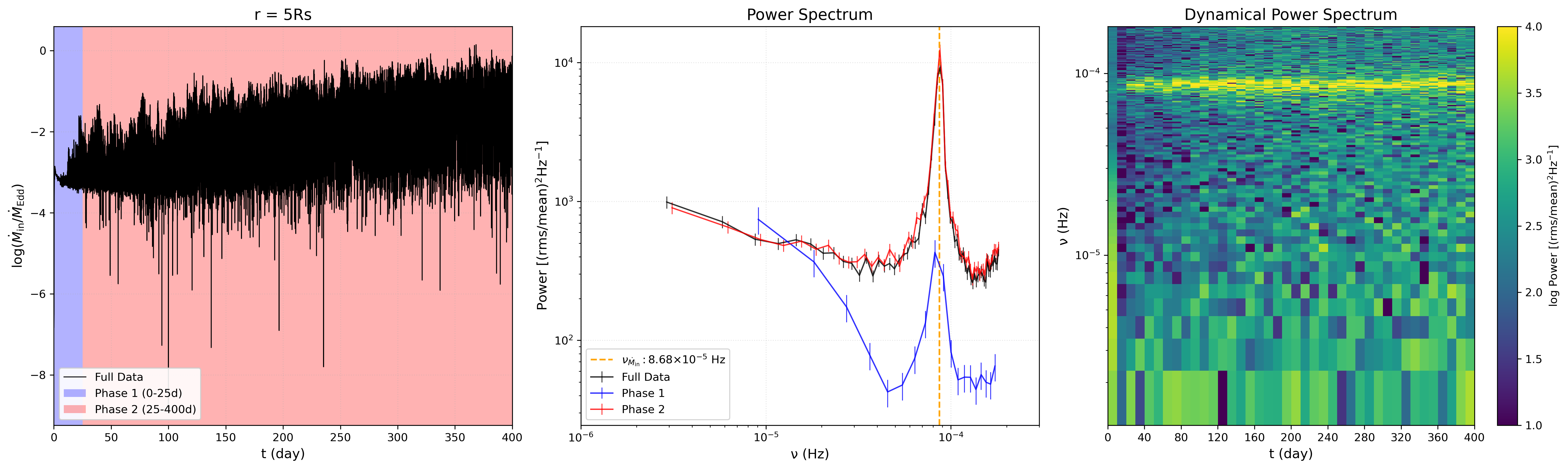}
    \hfill
    \includegraphics[width=0.49\textwidth]{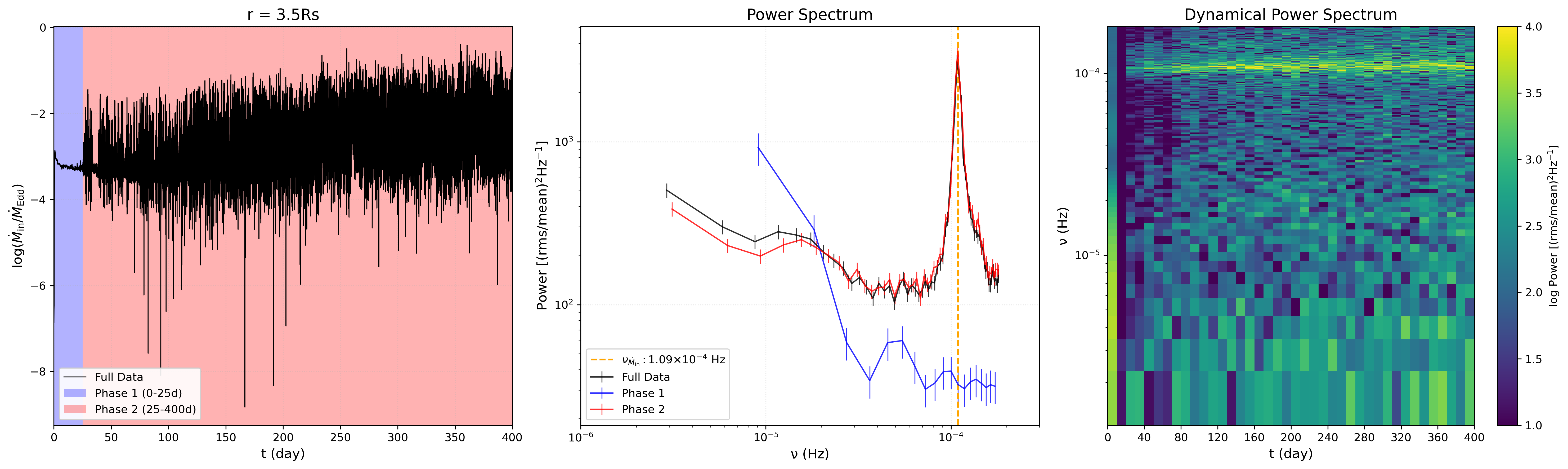}
    \hspace{0.001\textwidth}
    \includegraphics[width=0.49\textwidth]{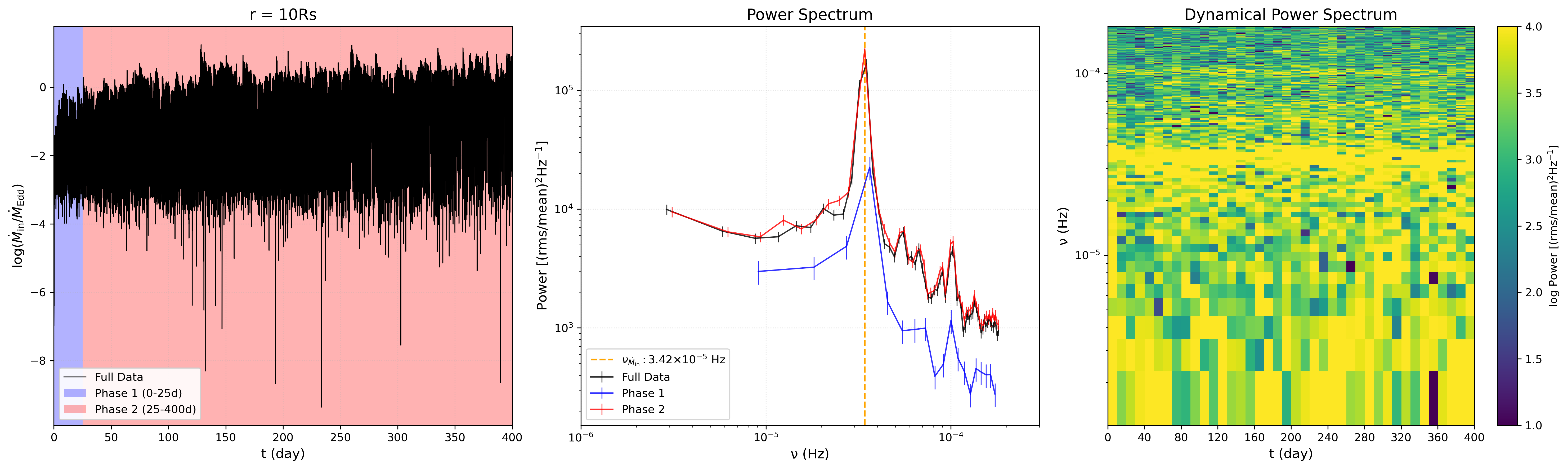}
    \hfill
    \includegraphics[width=0.49\textwidth]{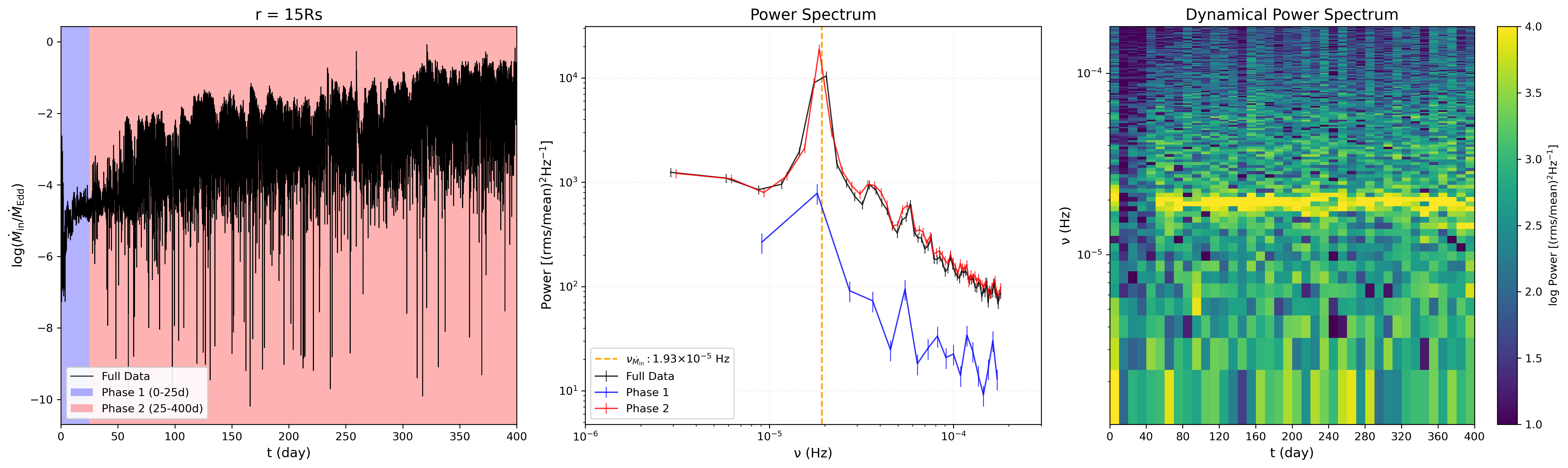}
    \hspace{0.001\textwidth}
    \caption{Evolution of the scaled mass inflow rate $\dot M_{\rm in}/\dot M_{\rm Edd}$, the corresponding power spectrum and the dynamical power spectrum with mass injection rate $\dot M_{\rm inject} = 0.1\dot M_{\rm Edd}$. Left side: results for $r=2R_{\rm S}$, $r=3R_{\rm S}$ and $r=3.5R_{\rm S}$ (from top to bottom). Right side: results for $r=3.8R_{\rm S}$, $r=5R_{\rm S}$ and $r=10R_{\rm S}$ (from top to bottom). Bottom side: results for $r=15R_{\rm S}$. For each figure at specific $r$, 
    left panel shows $\dot M_{\rm in}/\dot M_{\rm Edd}$ as a function of $t$  respectively. The purple region and the red region refer to phase 1 ($0-25$) day and phase 2 ($25-400$) day respectively; middle panel shows power spectrum calculated with FFT for the full data, phase I data and phase 2 data  respectively; right panel shows dynamical power spectrum calculated with FFT  respectively.}
    \label{fig:10}
\end{figure*}

\begin{figure*}
    \centering
    \includegraphics[width=0.49\textwidth]{figures/1M7-2Mdot/2Rs.png}
    \hspace{0.001\textwidth}
    \includegraphics[width=0.49\textwidth]{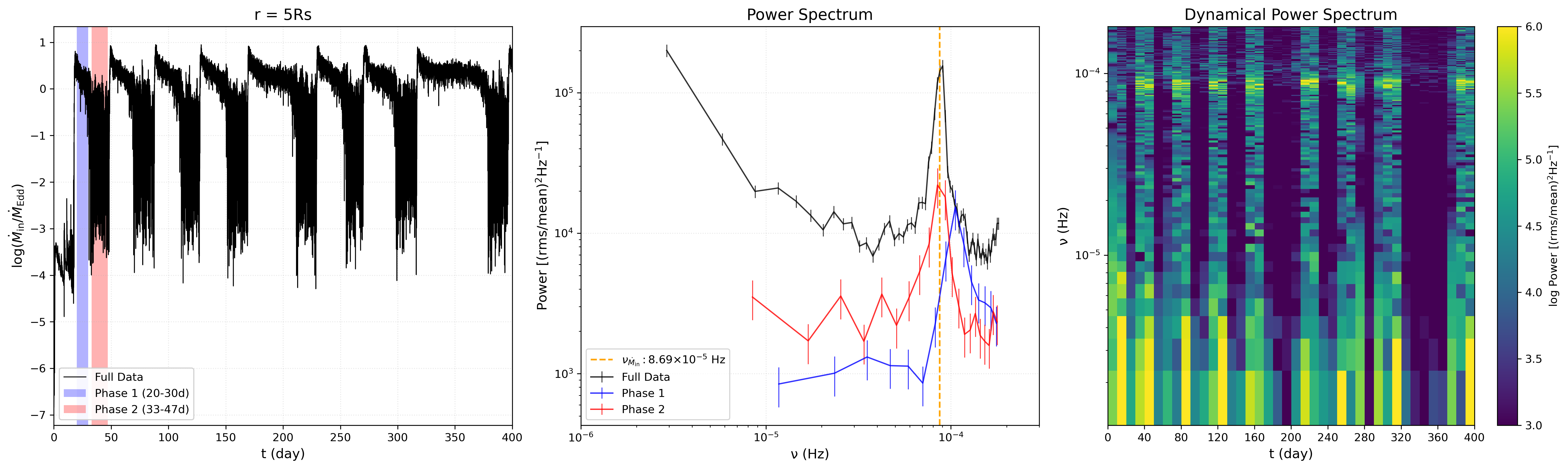}
    \hfill
    \includegraphics[width=0.49\textwidth]{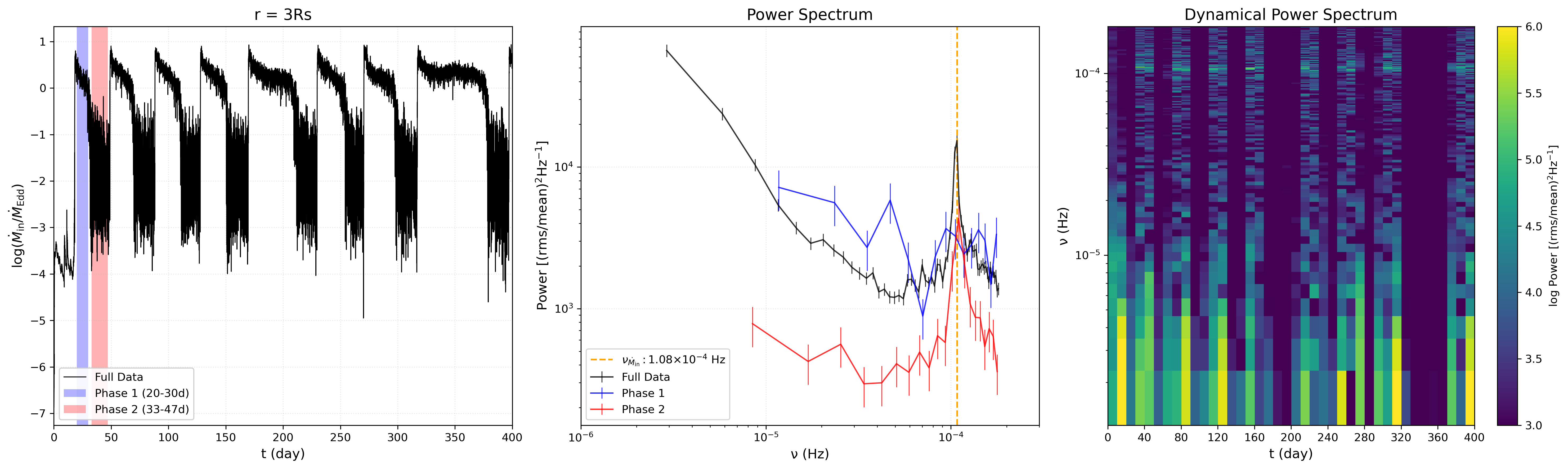}
    \hspace{0.001\textwidth}
    \includegraphics[width=0.49\textwidth]{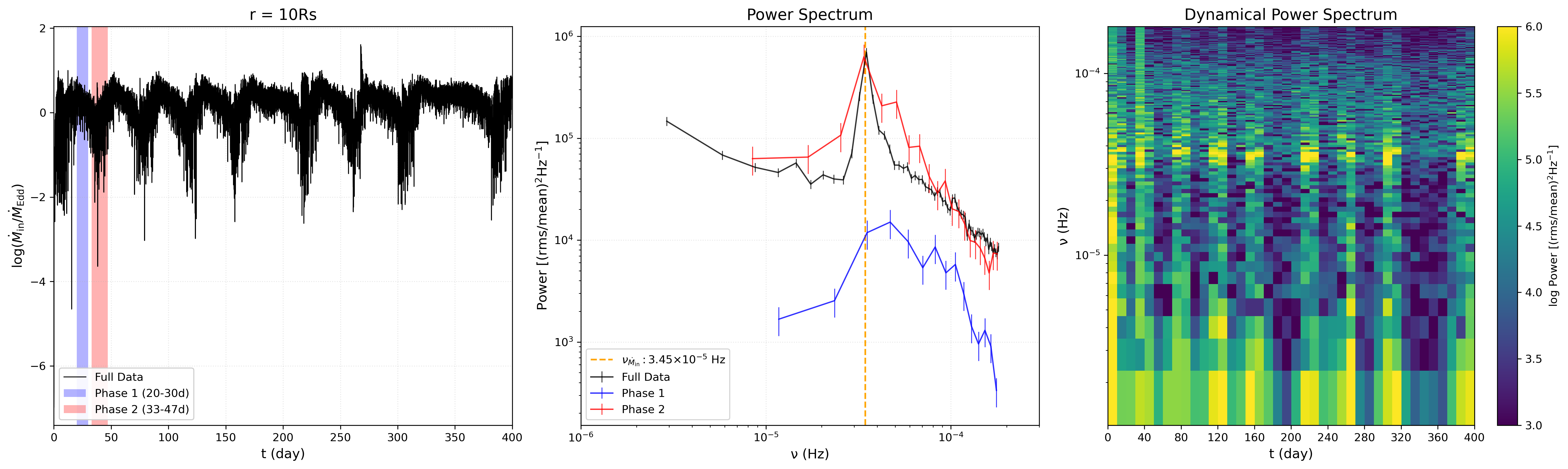}
    \hfill
    \includegraphics[width=0.49\textwidth]{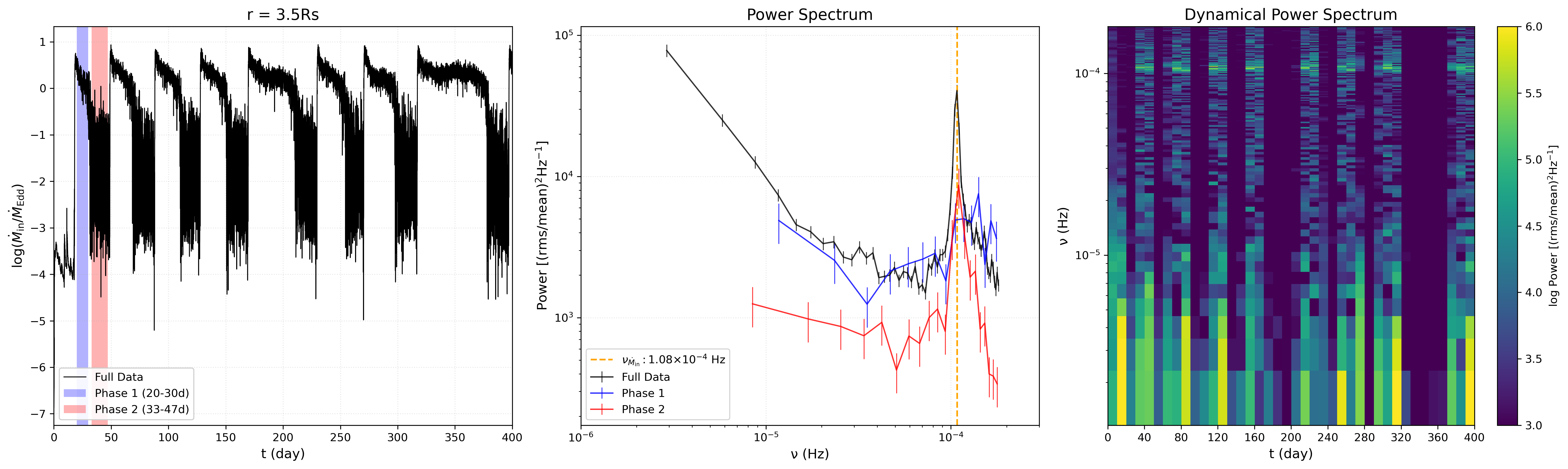}
    \hspace{0.001\textwidth}
    \includegraphics[width=0.49\textwidth]{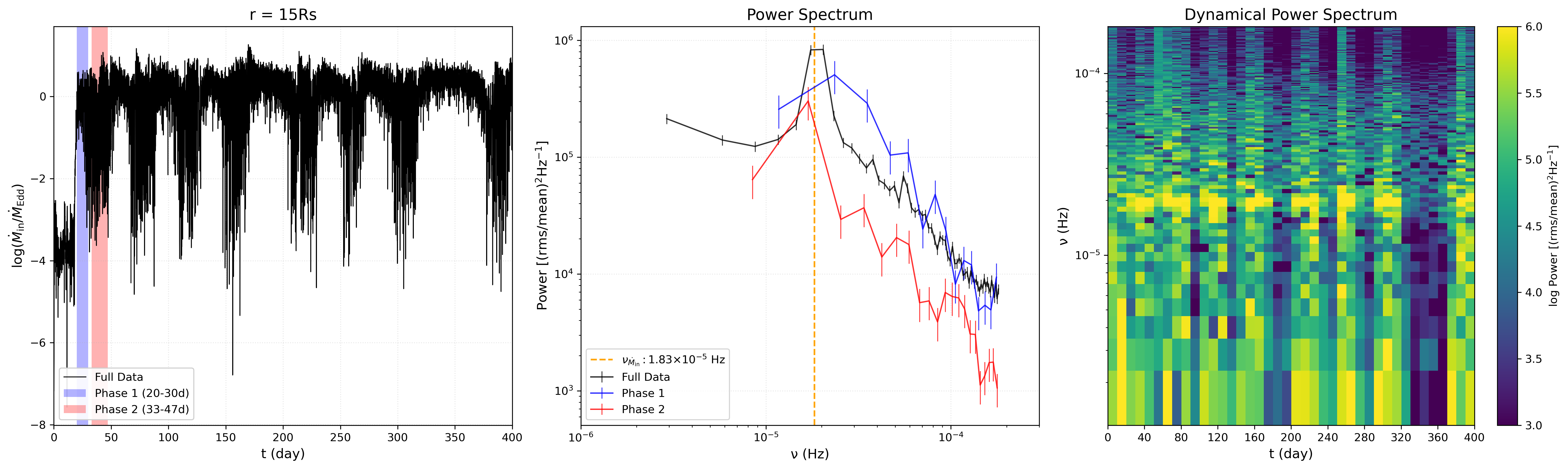}
    \hfill
    \includegraphics[width=0.49\textwidth]{figures/1M7-2Mdot/3.8Rs.png}
    \hspace{0.001\textwidth}
    \includegraphics[width=0.49\textwidth]{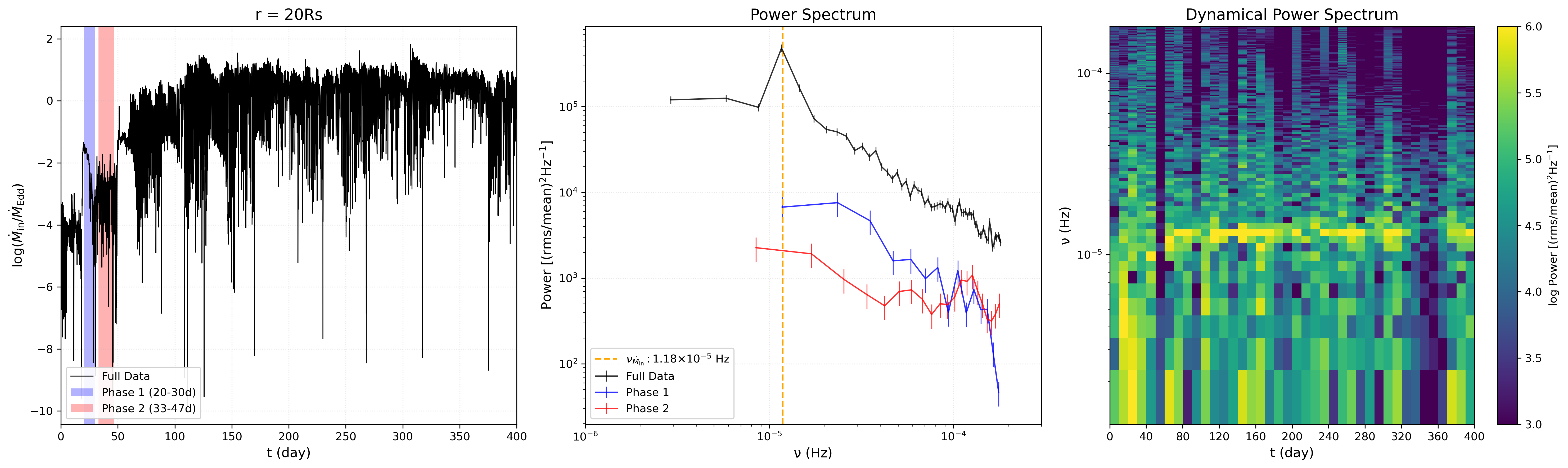}
    \caption{Evolution of the scaled mass inflow rate $\dot M_{\rm in}/\dot M_{\rm Edd}$, the corresponding power spectrum and the dynamical power spectrum with mass injection rate $\dot M_{\rm inject} = 2\dot M_{\rm Edd}$. Left side: results for $r=2R_{\rm S}$, $r=3R_{\rm S}$, $r=3.5R_{\rm S}$ and $r=3.8R_{\rm S}$ (from top to bottom). Right side: results for $r=5R_{\rm S}$, $r=10R_{\rm S}$, $r=15R_{\rm S}$ and $r=20R_{\rm S}$ (from top to bottom). For each figure at specific $r$, 
    left panel shows $\dot M_{\rm in}/\dot M_{\rm Edd}$ as a function of $t$  respectively. The purple region and the red region refer to phase 1 ($20-30$) day and phase 2 ($33-47$) day respectively; middle panel shows power spectrum calculated with FFT for the full data, phase I data and phase 2 data  respectively; right panel shows dynamical power spectrum calculated with FFT  respectively.}
    \label{fig:11}
\end{figure*}

\clearpage
\newpage

\section{Effect of different gravitational field forms}\label{sec:Appendix B}

\begin{figure*}
    \centering
    \includegraphics[width=0.7\textwidth]{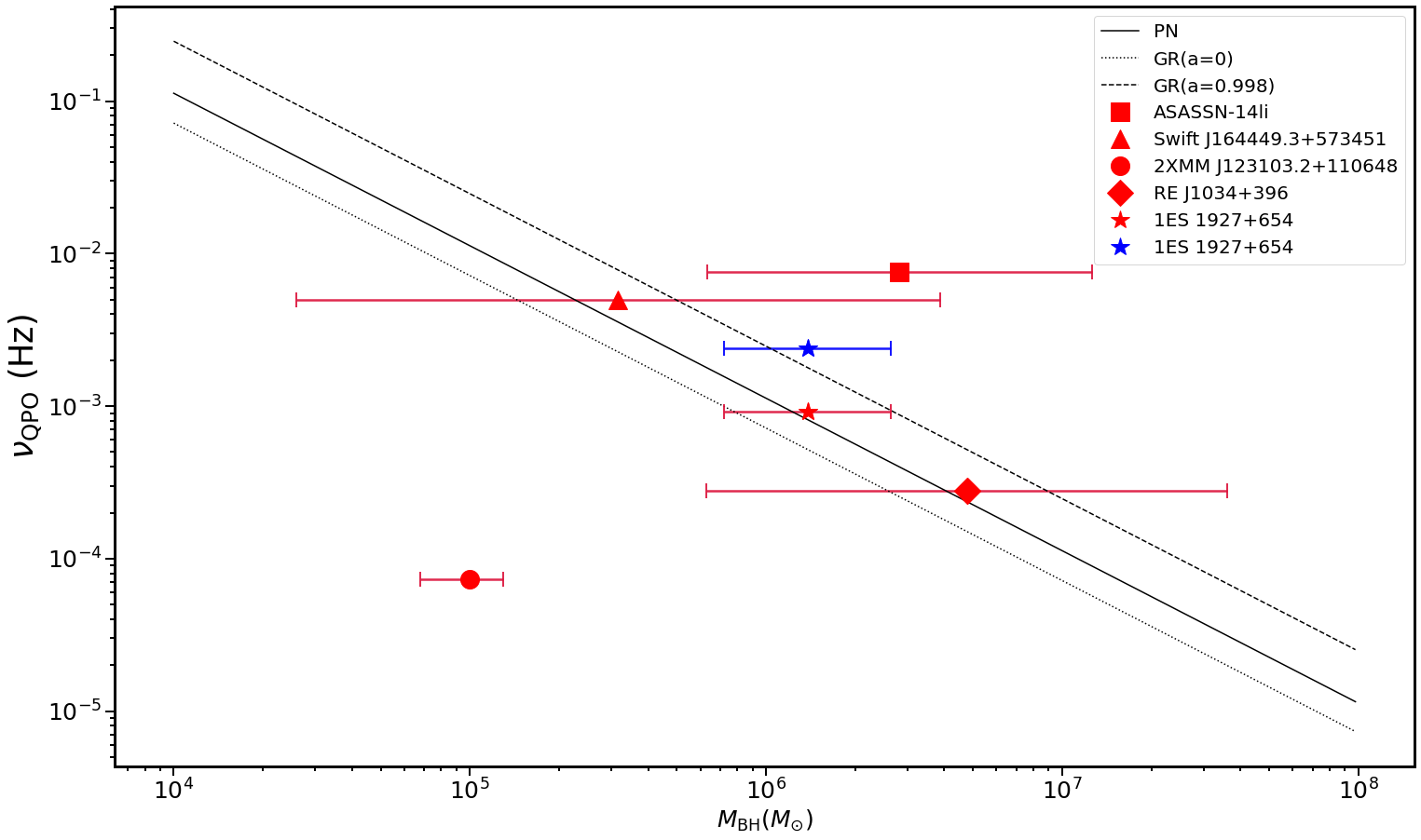}
    \caption{Relation between the observed QPO frequency $\nu_{\rm QPO}$ and the BH mass $M_{\rm BH}$ for five sources, including two AGNs, i.e, RE J1034+396, 1ES 1927+654, as well as three TDEs, i.e., Swift J164449.3+573451, ASASSN-14li and 2XMM J123103.2+110648 (TDE candidate). The black solid line is the maximum radial epicyclic frequency $\nu_{\rm r, max}$ as a function of $M_{\rm BH}$ assuming pseudo-Newtonian potential, i.e equation (\ref{eq:5}). The dotted line and the dashed line are the maximum radial epicyclic frequency $\nu_{\rm r, max}$ as a function of $M_{\rm BH}$ assuming GR based on equation (\ref{eq:6}) and (\ref{eq:7}) for taking $a=0$ and $a=0.998$ respectively.}
    \label{fig:12}
\end{figure*}

\clearpage
\newpage

\section{FFT parameters and results} \label{sec:Appendix C}

\clearpage
\newpage

\begin{table}
    \centering
	\caption{\large QPO frequencies for $\dot M_{\rm inject} = 0.01\dot M_{\rm Edd}$}
	\label{tab:1}
	\begin{minipage}{0.45\textwidth} 
    \centering
    \renewcommand{\arraystretch}{1.3}
    
	\begin{tabular}{cccc} 
		\hline
		\multirow{2}{*}{radius} & $\nu_{\dot M_{\rm in}}$(Hz) & $\nu_{\dot M_{\rm in}}$(Hz) & $\nu_{\dot M_{\rm in}}$(Hz)\\
		[3pt]
		& phase 1 ($0-41$d)& phase 2 ($41-400$d) & full data ($0-400$d) \\
		\hline
		$2R_{\rm S}$ & -- & $1.09\times 10^{-4}$ & $1.09\times 10^{-4}$\\
        \hline
        $3R_{\rm S}$ & -- & $1.09\times 10^{-4}$ & $1.09\times 10^{-4}$\\
        \hline
        $3.5R_{\rm S}$ & -- & $1.08\times 10^{-4}$ & $1.08\times 10^{-4}$\\
        \hline
        $3.8R_{\rm S}$ & -- & $1.08\times 10^{-4}$ & $1.08\times 10^{-4}$\\
        \hline
        $5R_{\rm S}$ & $8.68\times 10^{-5}$ & $8.68\times 10^{-5}$ & $8.68\times 10^{-5}$\\
        \hline
        $10R_{\rm S}$ & $3.41\times 10^{-5}$ & $3.41\times 10^{-5}$ & $3.41\times 10^{-5}$\\
        \hline
        $15R_{\rm S}$ & -- & $1.90\times 10^{-5}$ & $1.90\times 10^{-5}$\\
        \hline
        $20R_{\rm S}$ & -- & $1.27\times 10^{-5}$ & $1.27\times 10^{-5}$\\
        \hline
	\end{tabular}
	\end{minipage}
	\begin{minipage}{0.45\textwidth} 
    \vspace{2mm}
    \footnotesize \textit{Note.} QPO frequencies $\nu_{\dot M_{\rm in}}$ for selected data phases (phase 1, 2 and full data) of $\dot M_{\rm inject} = 0.01\dot M_{\rm Edd}$, including radius at $r=2,3,3.5,3.8,5,10,15,20R_{\rm S}$. Time resolution for the simulation data is $\Delta t = 2746.8s$. The segment numbers for using \textit{AveragedPowerspectrum} package in \textit{Stingray} to doing FFT are $20,100,100$ for phase 1, 2 and full data.
    \end{minipage}
\end{table}

\begin{table}
    \centering
	\caption{\large QPO frequencies for $\dot M_{\rm inject} = 0.1\dot M_{\rm Edd}$}
	\label{tab:2}
	\begin{minipage}{0.45\textwidth} 
    \centering
    \renewcommand{\arraystretch}{1.3}
    
	\begin{tabular}{cccc} 
		\hline
		\multirow{2}{*}{radius} & $\nu_{\dot M_{\rm in}}$(Hz) & $\nu_{\dot M_{\rm in}}$(Hz) & $\nu_{\dot M_{\rm in}}$(Hz)\\
		[3pt]
		& phase 1 ($0-25$d)& phase 2 ($25-400$d) & full data ($0-400$d) \\
		\hline
		$2R_{\rm S}$ & -- & $1.07\times 10^{-4}$ & $1.07\times 10^{-4}$\\
        \hline
        $3R_{\rm S}$ & -- & $1.07\times 10^{-4}$ & $1.07\times 10^{-4}$\\
        \hline
        $3.5R_{\rm S}$ & -- & $1.09\times 10^{-4}$ & $1.09\times 10^{-4}$\\
        \hline
        $3.8R_{\rm S}$ & -- & $1.09\times 10^{-4}$ & $1.09\times 10^{-4}$\\
        \hline
        $5R_{\rm S}$ & $8.68\times 10^{-5}$ & $8.68\times 10^{-5}$ & $8.68\times 10^{-5}$\\
        \hline
        $10R_{\rm S}$ & $3.42\times 10^{-5}$ & $3.42\times 10^{-5}$ & $3.42\times 10^{-5}$\\
        \hline
        $15R_{\rm S}$ & $1.93\times 10^{-5}$ & $1.93\times 10^{-5}$ & $1.93\times 10^{-5}$\\
        \hline
	\end{tabular}
	\end{minipage}
	\begin{minipage}{0.45\textwidth} 
    \vspace{2mm}
    \footnotesize \textit{Note.} QPO frequencies $\nu_{\dot M_{\rm in}}$ for selected data phases (phase 1, 2 and full data) of $\dot M_{\rm inject} = 0.1\dot M_{\rm Edd}$, including radius at $r=2,3,3.5,3.8,5,10,15R_{\rm S}$. Time resolution for the simulation data is $\Delta t = 2746.8s$. The segment numbers for using \textit{AveragedPowerspectrum} package in \textit{Stingray} to doing FFT are $20,100,100$ for phase 1, 2 and full data.
    \end{minipage}
\end{table}

\begin{table}
    \centering
	\caption{\large QPO frequencies for $\dot M_{\rm inject} = 2\dot M_{\rm Edd}$}
	\label{tab:3}
	\begin{minipage}{0.45\textwidth} 
    \centering
    \renewcommand{\arraystretch}{1.3}
    
	\begin{tabular}{cccc} 
		\hline
		\multirow{2}{*}{radius} & $\nu_{\dot M_{\rm in}}$(Hz) & $\nu_{\dot M_{\rm in}}$(Hz) & $\nu_{\dot M_{\rm in}}$(Hz)\\
	    [3pt]
		& phase 1 ($20-30$d)& phase 2 ($33-47$d) & full data ($0-400$d) \\
		\hline
		$2R_{\rm S}$ & -- & $1.08\times 10^{-4}$ & $1.08\times 10^{-4}$\\
        \hline
        $3R_{\rm S}$ & -- & $1.08\times 10^{-4}$ & $1.08\times 10^{-4}$\\
        \hline
        $3.5R_{\rm S}$ & -- & $1.08\times 10^{-4}$ & $1.08\times 10^{-4}$\\
        \hline
        $3.8R_{\rm S}$ & $1.18\times 10^{-4}$ & $1.08\times 10^{-4}$ & $1.08\times 10^{-4}$\\
        \hline
        $5R_{\rm S}$ & $1.07\times 10^{-4}$ & $8.69\times 10^{-5}$ & $8.69\times 10^{-5}$\\
        \hline
        $10R_{\rm S}$ & -- & $3.45\times 10^{-5}$ & $3.45\times 10^{-5}$\\
        \hline
        $15R_{\rm S}$ & -- & $1.83\times 10^{-5}$ & $1.83\times 10^{-5}$\\
        \hline
        $20R_{\rm S}$ & -- & -- & $1.18\times 10^{-5}$\\
        \hline
	\end{tabular}
	\end{minipage}
	\begin{minipage}{0.45\textwidth} 
    \vspace{2mm}
    \footnotesize \textit{Note.} QPO frequencies $\nu_{\dot M_{\rm in}}$ for selected data phases (phase 1, 2 and full data) of $\dot M_{\rm inject} = 2\dot M_{\rm Edd}$, including radius at $r=2,3,3.5,3.8,5,10,15,20R_{\rm S}$. Time resolution for the simulation data is $\Delta t = 2746.8s$. The segment numbers for using \textit{AveragedPowerspectrum} package in \textit{Stingray} to doing FFT are $10,10,100$ for phase 1, 2 and full data.
    \end{minipage}
\end{table}

\bsp	
\label{lastpage}
\end{document}